\documentclass[twocolumn,showpacs,preprintnumbers,amsmath,amssymb]{revtex4}

\usepackage{graphicx}
\usepackage{dcolumn}
\usepackage{bm}

\begin{document}

\title{Effective low energy theory for surface Andreev bound states of superconducting phases in generalized Bernevig-Hughes-Zhang models}

\author{Lei Hao$^{1}$ and Ting-Kuo Lee$^{2}$}
 \address{$^1$Department of Physics, Southeast University, Nanjing 210096, China   \notag \\
          $^2$Institute of Physics, Academia Sinica, Nankang, Taipei 11529, Taiwan}

\date{\today}

\begin{abstract}
A method for constructing the low energy effective models for pairings in the generalized Bernevig-Hughes-Zhang model for materials like Bi$_{2}$Se$_{3}$ is proposed. Pairings in this two-orbital model are identified with those familiar in one-orbital models, enabling a unified understanding. The theory provides an easy way to understand the topological nature of the superconducting state that is not directly related to the topological order in the normal state but due to subtle coupling among the degrees of freedom. Furthermore this approach shows a simple way to characterize the anisotropic nature of surface Andreev bound states (SABSs). In particular, we have identified the conditions to have a surprising new result of having two pairs of SABSs. It also leads to a conclusion that SABSs always connect with the topological surface states if the latter are well defined at the chemical potential.
\end{abstract}


\maketitle

\section{\label{sec:Introduction}Introduction}
Time reversal invariant (TRI) topological superconductor (TSC), which was proposed as a generalization of the TRI topological insulator (TI), has become a research focus of condensed matter physics.\cite{schnyder08,qi09,qi1081,qi10,sato09,sato10,teo10,roy08,kitaev09,fu10,hao11,sasaki11,hsieh12,basis,qc,yamakage12,hor10,wray10,zhang11pa,zhang11pb,kirshenbaum13,koren11,kirzhner12,chen12,hao13,levy13,qirmp,sasaki12,novak13,tanaka12,nakosai12,michaeli12,zhang13l} One promising candidate of TRI TSC is Cu$_{x}$Bi$_{2}$Se$_{3}$\cite{hor10,wray10}, which is also the first superconductor (SC) realized from a three dimensional (3D) TI. Intensive subsequent searches have added to the candidate list of TRI TSC the Bi$_{2}$X$_{3}$ (X is Se or Te) under high pressure\cite{zhang11pa,zhang11pb,kirshenbaum13} and the In-doped SnTe\cite{sasaki12,novak13}. A common feature of these SCs is that, their normal phases are described by the same kind of low energy effective model\cite{sasaki12,fu10}, which we call here as the generalized Bernevig-Hughes-Zhang (BHZ) model.\cite{bernevig06}

One of the most remarkable features of TRI TSC is that they support surface Andreev bound states (SABSs) known as Majorana fermions, which are promising candidates for realizing the fault-tolerant topological quantum computations.\cite{schnyder08,qi09,roy08,sato09,qirmp,wilczeketal} The SABSs are also essential to various transport properties, such as the tunneling spectroscopy.\cite{yamakage12} Several candidate pairings for Cu$_{x}$Bi$_{2}$Se$_{3}$ and similar materials are known to support SABSs.\cite{fu10,hao11,sasaki11,hao13} An interesting feature of the SABSs for some topological nontrivial pairings of superconducting TIs (STIs) is that, they connect continuously to the topological surface states (TSSs) inherited from the normal phase (henceforth called SABS-TSS connection).\cite{hao11,sasaki11,hsieh12} While this feature was explained as an accidental crossing between the SABSs and the TSSs\cite{hao11}, and also in terms of mirror symmetry breaking by the pairing\cite{hsieh12,yamakage12}, whether or not it occurs for all pairings that support SABSs is still unclear.\cite{hao11} Besides, it is not clear from the existing works whether new kinds of SABSs can be expected for pairings realized in the generalized BHZ model. Since the generalized BHZ model is a two-orbital model, it is not easy to extract physics by working with the full model directly. A simple picture for the pairings in these materials that can give a fast and reliable answer for the existence and property of the SABSs is thus highly desired.

In this work, we propose a simple and unified way to understand the low energy properties of these exotic SCs, the STIs in particular. Exploiting the close analogy between TIs and single orbital TRI TSCs, we show that if a pairing in the two-orbital generalized BHZ model is topological nontrivial, the full model can be reduced to low energy effective models corresponding to conventional $p$ wave triplet SCs, all well-known in the study of $^{3}$He.\cite{salomaa87} The normal state Z$_{2}$ topological order \cite{qipt,kane05,zhang09} is found \emph{not} essential for a certain pairing to be topological. However, for Z$_{2}$ nontrivial normal state, the topological pairing can own one or two Kramers' pairs of SABSs, depending on the relative position of the chemical potential with respect to the bands. In addition, we show that the SABS-TSS connection is a universal character of STIs. Thus, this new approach not only provides a unified picture for pairings in two-orbital BHZ type models and in one-orbital models but also provides a very useful and intuitive way to understand the low energy properties (e.g., SABSs) of STIs and related SCs in terms of well-known prototypes.

In the rest of the paper, we first present the model, the main results and related discussions in Sec.II. Then we give a short summary in Sec.III. The technical details are provided in the Appendices.

\section{\label{sec:result} results and discussions}

Bi$_{2}$Se$_{3}$ and other relevant materials are well described by a minimal two-orbital $\mathbf{k\cdot p}$ model which can be regarded as a 3D generalization of the two dimensional (2D) BHZ model.\cite{sasaki12,bernevig06,zhang09,fu09,qipt,wang10,liu10} This model can be mapped to a 3D hexagonal lattice.\cite{hao11,sasaki11,wang10,hao13} Define the basis vector for a wave vector $\mathbf{k}$ in the hexagonal BZ as $\phi_{\mathbf{k}}^{\dagger}=[a_{\mathbf{k}\uparrow}^{\dagger}
,a_{\mathbf{k}\downarrow}^{\dagger},b_{\mathbf{k}\uparrow}^{\dagger},b_{\mathbf{k}\downarrow}^{\dagger}]$,
in which $a$ and $b$ operators correspond to the two orbitals, the Hamiltonian matrix is written as\cite{liu10,wang10,fu09,hao11,hao13}
\begin{eqnarray}
h(\mathbf{k})&=&\epsilon(\mathbf{k})\sigma_{0}\otimes s_{0}+m(\mathbf{k})\sigma_{3}\otimes s_{0}   \notag \\
&&+\sigma_{1}\otimes[c_{y}(\mathbf{k})s_{1}-c_{x}(\mathbf{k})s_{2}]+c_{z}(\mathbf{k})\sigma_{2}\otimes s_{0},
\end{eqnarray}
where $s_{\alpha}$ and $\sigma_{\alpha}$ ($\alpha$=1,2,3) are Pauli matrices in the spin and orbital subspaces, respectively. $s_{0}$ and $\sigma_{0}$ are 2 by 2 unit matrices. The inversion operator is $P=\sigma_{3}\otimes s_{0}$.\cite{hao11} In this orbital convention, $a$ and $b$ have even and odd parities, respectively.\cite{zhang09,liu10} A slightly different model, with the $c_{z}(\mathbf{k})\sigma_{2}\otimes s_{0}$ term in Eq.(1) replaced by $c_{z}(\mathbf{k})\sigma_{1}\otimes s_{3}$, is also widely used in the literature.\cite{zhang09,hao11,qipt} This modified model (henceforth called modified Eq.(1)), though does not respect the mirror symmetry of actual materials\cite{hsieh12}, does describe a TI.\cite{zhang09,liu10,hao11} We would consider mostly Eq.(1) and sometimes also use the modified model in our analysis. Take lattice parameters as length units, we have $m(\mathbf{k})=m_{0}+2m_{1}(3-2\cos\frac{\sqrt{3}}{2}k_{x}\cos\frac{1}{2}k_{y}-\cos k_{y})+2m_{2}(1-\cos k_{z})$, where $m_{1}m_{2}>0$. $\epsilon(\mathbf{k})$ is obtained from $m(\mathbf{k})$ by substituting $\epsilon_{\alpha}$ for $m_{\alpha}$ and describes a topologically trivial band shift. $c_{x}(\mathbf{k})=\frac{2}{\sqrt{3}}A\sin\frac{\sqrt{3}}{2}k_{x}\cos\frac{1}{2}k_{y}$, $c_{y}(\mathbf{k})=\frac{2}{3}A(\cos\frac{\sqrt{3}}{2}k_{x}\sin\frac{1}{2}k_{y}+\sin k_{y})$, and $c_{z}(\mathbf{k})=B\sin k_{z}$. Two ingredients are known to be essential to make the (generalized) BHZ model a description of TIs.\cite{qipt,qirmp} The first is the existence of couplings between the two orbitals, which must be odd functions of $\mathbf{k}$ since $a$ and $b$ are of opposite parity and the materials have inversion symmetry. This is embodied in the $c_{\alpha}(\mathbf{k})$ ($\alpha=x,y,z$) terms originating from spin-orbit interaction.\cite{zhang09,liu10} The other is the existence of \emph{band inversion} (BI), in which the ordering of the two orbitals are inverted somewhere in the BZ, satisfied when $m_{0}m_{1}<0$ and $|m_{0}/m_{\alpha}|$ ($\alpha$=1,2) are not too large.\cite{bernevig06,zhang09,qipt,hao11}

We now review and emphasize the similarity between TIs, focusing on the generalized BHZ model, and one-orbital TRI TSCs.\cite{qi09,qi1081,qi10,sato09,qirmp,schnyder08,roy08,sato10} Historically, TRI TSCs were found decades ago (e.g., the BW phase of $^{3}$He\cite{balian63}) and were reformulated recently in close analogy to TIs.\cite{qi09,schnyder08,roy08} The analogy is rooted in the similarity between the Bogoliubov-de Gennes (BdG) Hamiltonian for a SC and the Hamiltonian for a band insulator. To establish a formal equivalence in terms of Eq.(1) for the generalized BHZ model, let us set the topologically irrelevant $\epsilon(\mathbf{k})$ term to zero hereafter in this work (See Appendix D for an analysis of this approximation). Then make a rotation of $\frac{\pi}{2}$ round the $x$ axis in the spin subspace by $U=\sigma_{0}\otimes e^{\frac{i\pi}{4}s_{1}}$, and a further substitution of $-c_{y}(\mathbf{k})$ for $c_{z}(\mathbf{k})$ and $c_{z}(\mathbf{k})$ for $c_{y}(\mathbf{k})$ which amounts to a rotation of $\frac{\pi}{2}$ round $k_{x}$ axis in the $\mathbf{k}$ space, Eq.(1) becomes $m(\mathbf{k})\sigma_{3}\otimes s_{0}-c_{x}(\mathbf{k})\sigma_{1}\otimes s_{3}-c_{y}(\mathbf{k})\sigma_{2}\otimes s_{0}+c_{z}(\mathbf{k})\sigma_{1}\otimes s_{1}$. Now, reinterpret the $a$ and $b$ orbitals as the particle and hole bands in a single band SC, and noting that $c_{\alpha}(\mathbf{k})\sim k_{\alpha}$ ($\alpha=x,y,z$) for $\mathbf{k}\simeq\mathbf{0}$, the above model is \emph{exactly} the BdG Hamiltonian of the BW state, up to an anisotropy between the $k_{x}k_{y}$ plane and the $k_{z}$ direction.\cite{balian63,salomaa87,qi1081} For the modified Eq.(1), if we first change the basis for $b$ orbital to $[b^{\dagger}_{\mathbf{k}\eta},b^{\dagger}_{\mathbf{k}\nu}]= [b^{\dagger}_{\mathbf{k}\uparrow},b^{\dagger}_{\mathbf{k}\downarrow}]is_{2}$ and then make the substitution of $c_{x}(\mathbf{k})$ for $c_{y}(\mathbf{k})$ and $-c_{y}(\mathbf{k})$ for $c_{x}(\mathbf{k})$, we obtain the same model obtained above from Eq.(1). In this reinterpretation of (modified) Eq.(1), the BI condition amounts to the weak pairing condition, that is the chemical potential (played by $-m_{0}$) must cross the conduction or valence band\cite{read00}, and the linear in $\mathbf{k}$ couplings between $a$ and $b$ orbitals amount to $p$ wave pairing between the equivalent particle and hole bands, which is triplet in nature in this new interpretation. By analogy with BI in TI, we call hereafter the weak pairing condition in terms of \emph{particle hole inversion} (PHI). So, it is clear from the above analogy that TRI TSC in single band (orbital) models can occur only for triplet pairings and when PHI occurs.

After establishing the analogy between TIs and TRI TSCs, we study pairings formed in the generalized BHZ model. Introducing the Nambu basis $\psi^{\dagger}_{\mathbf{k}}=[\phi^{\dagger}_{\mathbf{k}},\phi^{\text{T}}_{-\mathbf{k}}]$ and a pairing term $\underline{\Delta}(\mathbf{k})$, the BdG Hamiltonian is\cite{hao11}
\begin{equation}
H(\mathbf{k})=
\begin{pmatrix} h(\mathbf{k})-\mu\sigma_{0}\otimes s_{0} & \underline{\Delta}(\mathbf{k}) \\
-\underline{\Delta}^{\ast}(-\mathbf{k}) & \mu\sigma_{0}\otimes s_{0}-h^{\ast}(-\mathbf{k})
\end{pmatrix},
\end{equation}
where $\mu$ is the chemical potential.

\begin{figure}
\centering
\includegraphics[width=8cm,height=6.0cm,angle=0]{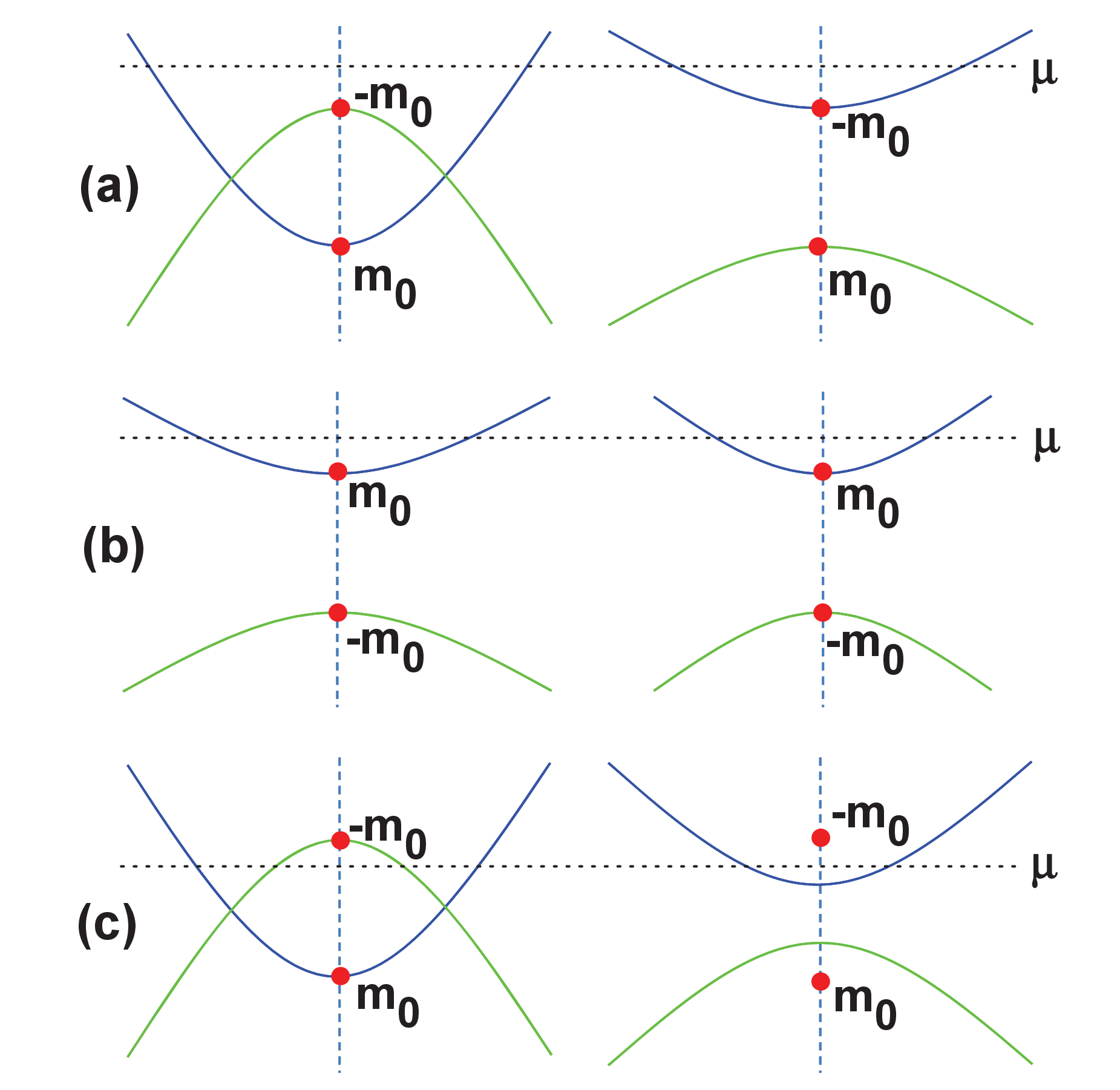}
\caption{Schematic drawing of three kinds of band structures that can occur in the generalized BHZ model, with the $\epsilon(\mathbf{k})$ term discarded. In each pair of figures in (a), (b) and (c), the left one shows the relative position of $m(\mathbf{k})$ (the upward parabola) and $-m(\mathbf{k})$ (the downward parabola), the right one shows the relative position of the bulk conduction band (the upward parabola) and bulk valence band (the downward parabola), in the normal phase. The horizontal dotted lines show positions of the chemical potential taken in this work. The two crossing points on each figure between the vertical dashed line and the two curves label the extremal points of $\pm m(\mathbf{k})$ or the bulk conduction and valence bands. In this work, the extremal points for the right figure of (c) correspond to $k_{x}=k_{y}=0$ and $k_{z}\ne0$, whereas they  correspond to $k_{x}=k_{y}=k_{z}=0$ for all other figures (See Appendix E).}
\end{figure}

We first study an interorbital triplet pairing which is topological nontrivial and have attracted the most attention to date, written as $\underline{\Delta}_{1}(\mathbf{k})=\Delta_{0}\sigma_{2}\otimes s_{1}$.\cite{fu10,hao11,sasaki11,hsieh12} As in all previous works, we first set $\mu$ far away from the TI gap center where the BI occurs.\cite{fu10,hao11,sasaki11,hsieh12} Without loss of generality, we first set the parameters as $m_{1}=m_{2}=0.5$, $A=1.5$, $B=1$, $m_{0}=-0.7$ and $\mu=0.9$.\cite{hao11} As shown in Fig.1(a) for a schematic drawing of this parameter set, only the $a$ orbital crosses $\mu$ and makes a PHI. Though the pairing operator contains the two orbitals on an equal footing, the $b$ orbital only sees a strong pairing field (in the naming convention of Read and Green\cite{read00}) and thus cannot hold a topological pairing.\cite{read00,qi09,qi1081} To be more specific, in terms of the Nambu description, the pairing introduces a hole orbital for both the $a$ and the $b$ particle orbitals. While the particle and hole $a$ orbitals meet and invert at $\mu$, the particle and hole $b$ orbitals are both far away from $\mu$, as shown in Fig. 1(a). Thus, the low energy quasiparticle excitations close to $\mu$ should be associated mostly with the $a$ orbital. Since we are interested only in low energy properties, in particular the SABSs, we could integrate out the high energy degrees of freedom corresponding to the $b$ orbital, and work with a low energy effective model associated \emph{mainly} with the $a$ orbital.

To simplify the deductions, we redefine the Nambu basis as $\psi^{\dagger}_{\mathbf{k}}=[\phi^{\dagger}_{\mathbf{k}a},\phi^{\dagger}_{\mathbf{k}b}]$, with $\phi^{\dagger}_{\mathbf{k}a}=[a^{\dagger}_{\mathbf{k}\uparrow},a_{-\mathbf{k}\uparrow},
a^{\dagger}_{\mathbf{k}\downarrow},a_{-\mathbf{k}\downarrow}]$ and $\phi^{\dagger}_{\mathbf{k}b}$ similarly. Labeled by the $a$ and $b$ orbitals, the BdG Hamiltonian $H(\mathbf{k})$ is written into block form. The upper-left diagonal block for the $a$ orbital is $h_{a}=[m(\mathbf{k})-\mu]s_{0}\otimes\tau_{3}$, the lower-right diagonal block for the $b$ orbital is $h_{b}=-[m(\mathbf{k})+\mu]s_{0}\otimes\tau_{3}$, the off-diagonal blocks are $h_{ab}=h^{\dagger}_{ba}=-ic_{z}(\mathbf{k})s_{0}\otimes\tau_{3}+c_{y}(\mathbf{k})s_{1}\otimes\tau_{0} -c_{x}(\mathbf{k})s_{2}\otimes\tau_{3}-i\Delta_{0}s_{1}\otimes\tau_{1}$. To get the low energy effective model within the $a$ orbital, we suppose $[u_{a},v_{b}]^{\text{T}}$ is an eigenfunction of $H(\mathbf{k})$ with energy $E$. Thus we have $h_{a}u_{a}+h_{ab}v_{b}=Eu_{a}$ and $h_{b}v_{b}+h_{ba}u_{a}=Ev_{b}$. Eliminate $v_{b}$, we get the eigen-equation for $a$ orbital as $[h_{a}+h_{ab}(E-h_{b})^{-1}h_{ba}]u_{a}=\tilde{h}_{a}u_{a}=Eu_{a}$. $\tilde{h}_{a}$ is the low energy effective model that we want.\cite{mccann06} Since we focus on states with $|E|\leq\Delta_{0}$, for small pairing amplitudes we could set $E=0$ in $\tilde{h}_{a}$ which amounts to neglecting small quantities proportional to $E/(m(\mathbf{k})+\mu)$ and their higher order terms (See Appendix C). Thus we get the low energy effective model within the $a$ orbital subspace as
\begin{eqnarray}
&&\tilde{h}_{a}\simeq h_{a}+h_{ab}(-h_{b})^{-1}h_{ba}   \notag \\
&&=[m-\mu+\frac{c^{2}_{x}+c^{2}_{y}+c^{2}_{z} -\Delta^{2}_{0}}{m+\mu}]s_{0}\otimes\tau_{3}   \notag \\
&&-\frac{2\Delta_{0}}{m+\mu} [c_{x}s_{3}\otimes\tau_{1}+c_{y}s_{0}\otimes\tau_{2}-c_{z}s_{1}\otimes\tau_{1}],
\end{eqnarray}
where we have made the $\mathbf{k}$ dependencies of the various terms implicit. A moment's reflection by taking into account the fact that $c_{\alpha}(\mathbf{k})\sim k_{\alpha}$ for $\mathbf{k}\simeq\mathbf{0}$ can convince one that, up to an anisotropy between the $k_{x}k_{y}$ plane and the $k_{z}$ direction, the effective triplet pairing in Eq.(3) describes the well-known BW phase.\cite{balian63,salomaa87}

According to the two criteria for identifying TSCs in a one-orbital model, we have to see whether a PHI still occurs in the effective model, that is whether or not $\frac{m^{2}+c^{2}_{x}+c^{2}_{y}+c^{2}_{z}-\mu^{2}-\Delta^{2}_{0}}{m+\mu}$ changes sign somewhere in the BZ, if $m-\mu$ changes sign in the BZ. We have found that, once $\Delta_{0}$ is small compared to $\sqrt{c^{2}_{x}+c^{2}_{y}+c^{2}_{z}}$ evaluated at wave vectors satisfying $m(\mathbf{k})=\mu$,
PHI always occurs in the effective model (See Appendix D). This is consistent with our assumption of small $\Delta_{0}$ and relevant experiments.\cite{hor10,wray10,zhang11pa,zhang11pb,kirshenbaum13} Recalling the identification to the BW phase, we know from previous works that, this pairing is fully gapped and is topologically nontrivial supporting SABSs on an arbitrary surface.\cite{fu10,hao11,sasaki11,roy08,qi09}

Furthermore, it is easy to see that PHI in the effective model occurs regardless of the value of $m_{0}$ and thus occurs even for $m_{0}m_{1}>0$ (no BI in the normal phase, see Fig.1(b) for a schematic illustration) which corresponds to a topologically trivial normal phase, such as Sb$_{2}$Se$_{3}$.\cite{zhang09} So, $\underline{\Delta}_{1}$ is TSC even if the normal phase is topologically trivial. To verify this result, we show in Figs.2(a) and 2(b) numerical results of the surface spectral functions for $m_{0}m_{1}<0$ and $m_{0}m_{1}>0$, respectively.\cite{hao11,hao13} In the small wave vector region, linearly dispersing SABSs appear in both two cases. The only qualitative difference appears close to the Fermi momentum, where the SABSs for a TI normal phase ($m_{0}m_{1}<0$) connects continuously to the TSSs. On account of this finding, ordinary semiconductors like Sb$_{2}$Se$_{3}$ with strong SOI are also promising candidates to find topological pairings.\cite{sasaki12}

\begin{figure}
\centering
\includegraphics[width=8.6cm,height=7.2cm,angle=0]{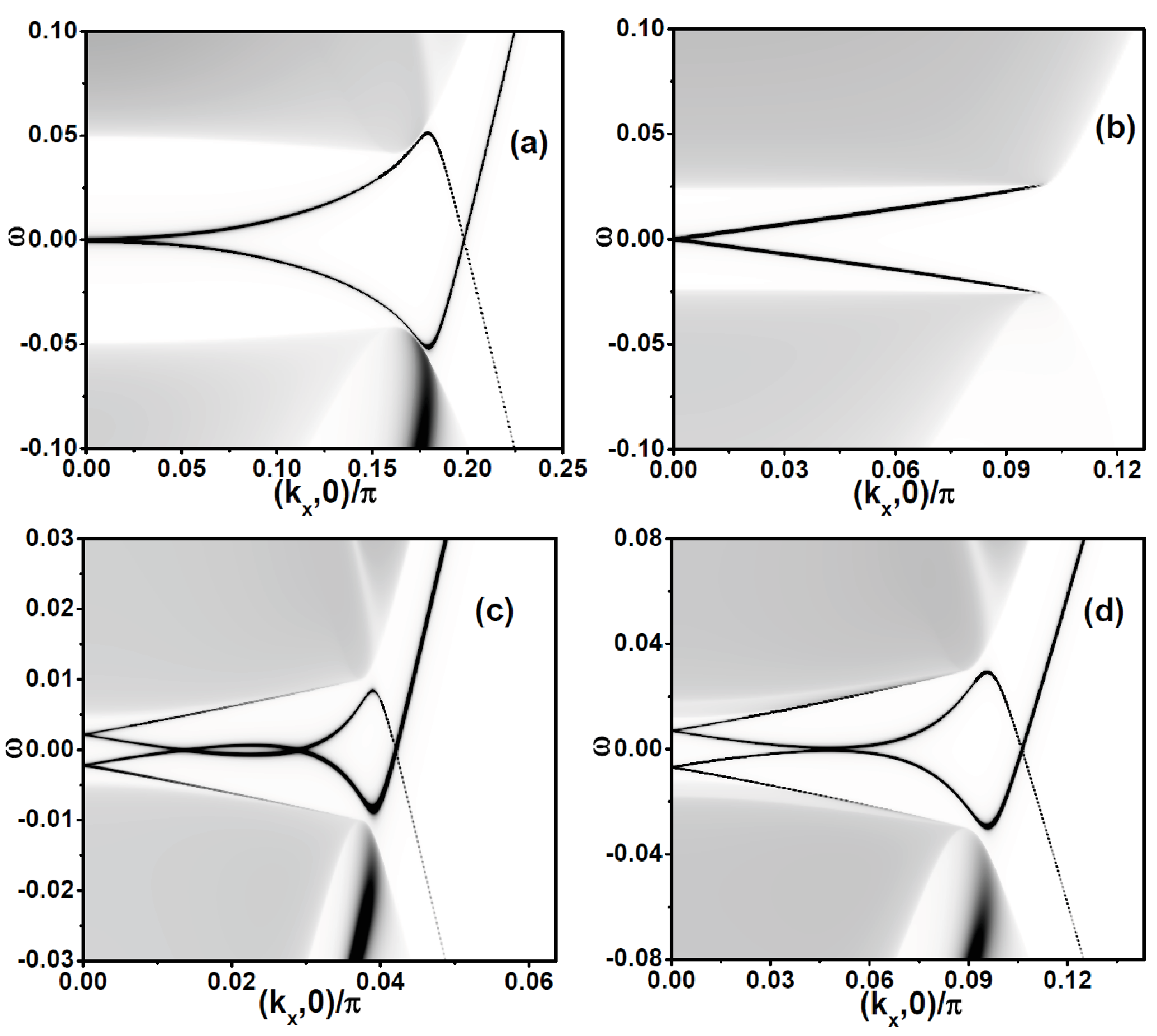}
\caption{Surface spectral functions for $\underline{\Delta}_{1}(\mathbf{k})$ in Eq.(1), for four parameter sets (neglecting the $\epsilon(\mathbf{k})$ term):
(a) $m_{1}=m_{2}=0.5$, $A=1.5$, $B=1$, $m_{0}=-0.7$, $\mu=0.9$, and $\Delta_{0}=0.05$;
(b) $m_{1}=m_{2}=0.5$, $A=1.5$, $B=1$, $m_{0}=0.7$, $\mu=0.9$, and $\Delta_{0}=0.05$;
(c) $m_{1}=m_{2}=0.5$, $A=1.5$, $B=0.1$, $m_{0}=-0.7$, $\mu=0.2$, and $\Delta_{0}=0.01$;
(d) $m_{1}=m_{2}=0.5$, $A=1.5$, $B=0.3$, $m_{0}=-0.7$, $\mu=0.5$, and $\Delta_{0}=0.03$.
The results are obtained in terms of standard iterative Green's function method.\cite{hao11,hao13} The darker the color, the larger the spectral weight.}
\end{figure}

Since the low energy effective models for the TSC phase within the $a$ orbital are essentially the same for $m_{0}m_{1}<0$ and $m_{0}m_{1}>0$, we are led to conclude that the parts of the two kinds of SABSs which have their origin in the TSC states are equivalent, and the additional feature for $m_{0}m_{1}<0$ in Fig.2(a) close to the Fermi momentum is a result of hybridization between the SABSs and the TSSs. The merge of SABS with TSSs is because on one hand the pairing $\underline{\Delta}_{1}(\mathbf{k})$ cannot open a gap within the TSSs and on the other hand the TSSs are well separated from the bulk states and thus well defined at $\mu$.\cite{hao11,yamakage12} Later in this work we would show that the SABS-TSS connection is likely a universal feature of STI.\cite{hao11,sasaki11,hsieh12}

Apart from the merge of SABS with TSSs, there is another important difference between $m_{0}m_{1}<0$ and $m_{0}m_{1}>0$. For $m_{0}m_{1}>0$ without of BI, PHI can only occur for a single orbital. But for $m_{0}m_{1}<0$ with BI, PHI can occur for both of the two orbitals if $A^{2}<3|m_{0}m_{1}|$ or $B^{2}<2|m_{0}m_{2}|$, and if we place $\mu$ properly ($\mu$ should cross one bulk band to ensure bulk pairing) within the region of $0<|\mu/m_{0}|<1$ (see Appendix E for more detailed explanations). See Fig.1(c) for a schematic illustration of the situation in the normal phase. In this case, the $a$ and $b$ orbitals both contribute nontrivially to the superconducting phase low energy excitations. For $|\mu|>0$, $m(\mathbf{k})-\mu$ and $-m(\mathbf{k})-\mu$ never change sign at the same wave vector (see the left figure of Fig.1(c) for an illustration of $\mu>0$). So, PHI for orbitals $a$ and $b$ occur at different portions of the BZ, which we denote as BZ$_{a}$ and BZ$_{b}$. For $\Delta_0\ll|\mu|$, BZ$_{a}$ and BZ$_{b}$ share no common wave vectors. We are thus justified to get two low energy effective models $\tilde{h}_{a}$ and $\tilde{h}_{b}$ in BZ$_{a}$ and BZ$_{b}$, respectively. Still consider $\underline{\Delta}_{1}$, $\tilde{h}_{a}$ is as shown in Eq.(3). $\tilde{h}_{b}$ is obtained from Eq.(3) by making the substitution of $\tilde{h}_{b}=\tilde{h}_{a}[m\rightarrow-m,\Delta_{0}\rightarrow-\Delta_{0},c_{z}\rightarrow-c_{z}]$. As a result, there would be two pairs of SABSs on each surface, originating separately from $\tilde{h}_{a}$ and $\tilde{h}_{b}$. The coupling between BZ$_{a}$ and BZ$_{b}$, upon the introduction of the surface, would induce hybridization between the two pairs of SABSs, so that the energies of the SABSs at the time reversal invariant momentum of the surface BZ (i.e., $k_{x}=k_{y}=0$) would be nonzero. As shown in Figs.2(c) and 2(d) are results for two typical parameter sets. The presence of two pairs of SABSs, and the nonzero excitation energies for $k_{x}=k_{y}=0$ are both obvious, which are easily understood with our low energy effective model approach. More discussions for the SABSs can be found in Appendix F. The finding of two pairs of SABSs is new in the present system and never shown before. Since in actual TIs, like Bi$_{2}$Se$_{3}$ and Bi$_{2}$Te$_{3}$, $A$ and $B$ are small enough and satisfy the conditions $A^{2}<3|m_{0}m_{1}|$ or $B^{2}<2|m_{0}m_{2}|$\cite{liu10,hao13}, it is quite possible to observe two pairs of SABS in experiments if the chemical potential is chosen properly.

So far we have shown with $\underline{\Delta}_{1}(\mathbf{k})$ as an example that the TSCs emerging from the generalized BHZ model can be understood clearly through the low energy effective models. This analysis is easily extended to other pairings. We would concentrate in the following on the cases in which BI occurs in the normal phase ($m_{0}m_{1}<0$) and only the $a$ orbital has a PHI ($\mu>-m_{0}>0$).

Two other triplet pairings are known to support nontrivial SABSs, which are $\underline{\Delta}_{2}(\mathbf{k})=\Delta_{0}\sigma_{2}\otimes s_{3}$ and $\underline{\Delta}_{3}(\mathbf{k})=i\Delta_{0}\sigma_{2}\otimes s_{0}$.\cite{fu10,hao11,sasaki11} Following the same analysis as above, we can get their low energy effective models. The effective pairings within the $a$ orbital subspace for $\underline{\Delta}_{2}(\mathbf{k})$ and $\underline{\Delta}_{3}(\mathbf{k})$ turn out to be equivalent to the planar phase of $^3$He\cite{salomaa87} in the $k_{x}k_{z}$ and $k_{y}k_{z}$ plane, respectively (see Appendix A).

An odd parity singlet pairing, $\underline{\Delta}_{4}(\mathbf{k})=i\Delta_{0}\sigma_{1}\otimes s_{2}$ is also of interest to us. Direct calculation (see Appendix A) shows that the low energy effective pairing within the $a$ orbital is also similar to the planar phase of $^3$He, but within the $k_{x}k_{y}$ plane.\cite{salomaa87} Since this effective pairing does not change sign across the $xy$ surface, it would not give SABSs lying in the $xy$ plane. But if a clean surface parallel to the $z$ axis can be prepared, this pairing should also give SABSs similar to the above two triplet pairings.\cite{hao11,sasaki11}

Recently, we have proposed a novel spin singlet pairing which gives flat-band SABSs, for Cu$_{x}$Bi$_{2}$Se$_{3}$ and Bi$_{2}$Te$_{3}$.\cite{hao13} In the present convention, it is defined as $\underline{\Delta}_{5}(\mathbf{k})=i\Delta_{0}\varphi_{\mathbf{k}}\sigma_{2}\otimes s_{2}$, where $\varphi_{\mathbf{k}}=\frac{2}{3}\sin\frac{k_{y}}{2}[\cos\frac{\sqrt{3}}{2}k_{x}+\cos\frac{1}{2}k_{x}]\sim k_{y}$ (or $\varphi_{\mathbf{k}}=\frac{2}{\sqrt{3}}\sin\frac{\sqrt{3}}{2}k_{x}\cos\frac{1}{2}k_{y}\sim k_{x}$). The low energy effective model for this pairing turns out to be $d_{yz}$ (or $d_{xz}$) wave singlet pairing (see Appendix A). In terms of this correspondence, the SABSs in this pairing is similar to the SABSs in cuprates.\cite{hu94}

It is now clear that pairing in the low energy effective model can be quite different from its original form. This originates from the inherent intricate couplings among the degrees of freedom in the model. Hence, a subtle change of these couplings would manifest itself in the change of the low-energy effective models, even if the topological (e.g., Z$_{2}$) nature of the normal phase remains unchanged. We illustrate this point by studying typical pairings formed in the modified Eq.(1).

First consider $\underline{\Delta}_{1}(\mathbf{k})$. The low energy effective model is easily obtained. Instead of a BW type pairing, the effective pairing in the low energy model is equivalent to the planar phase of $^{3}$He, within the $k_{x}k_{y}$ plane (see Appendix B).\cite{salomaa87} Since the order parameter does not change sign across the $xy$ plane, no SABSs are expected in direct contrast to the original model of Eq.(1).\cite{hao11} Now, consider the odd-parity singlet pairing $\underline{\Delta}_{4}(\mathbf{k})$. The low energy effective pairing within the $a$ orbital turns out to be equivalent to the BW phase by a proper redefinition of axes (see Appendix B), in comparison to the planar phase in the original model of Eq.(1).\cite{hao11} Thus we have seen a slight difference in the original model of Eq.(1) and the modified Eq.(1) will give very different SABSs. Thus specific coupling among the degrees of freedom determines the topological nature of a pairing.

We now come back to Eq.(1) and answer whether or not SABSs will always merge with TSSs in STIs that support SABSs, and TSSs are well separated from the bulk band states at $\mu$.\cite{hao11,sasaki11,hsieh12} This question can be answered by thinking reversely: If all pairings that can open a gap in the TSSs do not support SABSs on the $xy$ surface, then all pairings supporting SABSs on the $xy$ surface cannot open gap in the TSSs, hence the two different surface states will merge.

To test the above statement, we first construct the TRI pairing that can form in the TSSs.\cite{hao11} The TSSs on the $xy$ surface of a sample occupying the $z<0$ half space are obtained by solving the surface modes in terms of the continuum limit of Eq.(1).\cite{hao11,liu10} Denoting the surface states in terms of the basis $\phi_{\mathbf{k}}$ by making the $z$ dependency implicit, the two zero energy modes for $k_{x}=k_{y}=0$ are found to be $\eta_{1}=\frac{1}{\sqrt{2}}[1,0,-1,0]$ and $\eta_{2}=\frac{1}{\sqrt{2}}[0,1,0,-1]$. The effective model for the TSSs is\cite{hao11}
\begin{equation}
H_{eff}(\mathbf{k})=c_{x}(\mathbf{k})s_{2}-c_{y}(\mathbf{k})s_{1},
\end{equation}
where $\mathbf{k}$ is now defined in the surface BZ. Its two eigenvectors can be taken as $\eta_{\alpha}(\mathbf{k})=\frac{1}{\sqrt{2}}[\alpha\frac{c_{y}(\mathbf{k})+ic_{x}(\mathbf{k})} {\sqrt{c^{2}_{y}(\mathbf{k})+c^{2}_{x}(\mathbf{k})}},-1]^{\text{T}}$, with $\alpha=\pm$ and the corresponding eigenenergies $E_{\alpha}(\mathbf{k})=\alpha\sqrt{c^{2}_{x}(\mathbf{k})+c^{2}_{y}(\mathbf{k})}$. The creation operators for the eigenstates are thus $f^{\dagger}_{\mathbf{k},\alpha}=\frac{1}{2}[\alpha\frac{c_{y}(\mathbf{k})+ic_{x}(\mathbf{k})} {\sqrt{c^{2}_{y}(\mathbf{k})+c^{2}_{x}(\mathbf{k})}}(a^{\dagger}_{\mathbf{k}\uparrow}-b^{\dagger}_{\mathbf{k}\uparrow}) -(a^{\dagger}_{\mathbf{k}\downarrow}-b^{\dagger}_{\mathbf{k}\downarrow})]$.

Without loss of generality, consider pairing in the TSSs for $\mu>0$. The TRI pairing then must come in the form of $\frac{c_{y}(\mathbf{k})-ic_{x}(\mathbf{k})}{\sqrt{c^{2}_{y}(\mathbf{k})+c^{2}_{x}(\mathbf{k})}} f^{\dagger}_{\mathbf{k},+}f^{\dagger}_{-\mathbf{k},+}$.\cite{hao11} In terms of the $a$ and $b$ operators, this pairing can be decomposed into two singlet and two triplet pairings. One singlet pairing is $\underline{\Delta}_{6}(\mathbf{k})=i\Delta_{0}\sigma_{0}\otimes s_{2}$. For this pairing, the low energy effective pairing for the $a$ orbital is an $s$ wave singlet pairing with slight anisotropy and thus do not support SABSs (see Appendix A). Another singlet pairing is $\underline{\Delta}_{4}(\mathbf{k})$ studied earlier. This phase, though not completely topologically trivial, does not support SABSs on the $xy$ surface.

The two triplet pairings are both similar to the planar phase of $^{3}$He, including one intraorbital and one interorbital component. The interorbital component is $\underline{\Delta}_{7}(\mathbf{k})=\frac{\Delta_{0}}{A}\sigma_{1}\otimes[c_{y}(\mathbf{k})s_{1}-c_{x}(\mathbf{k})s_{2}]is_{2}$. The effective pairing within the $a$ orbital subspace is proportional to $[c^{2}_{x}(\mathbf{k})+c^{2}_{y}(\mathbf{k})]is_{2}$, an anisotropic $s$ wave pairing of approximately $s_{x^{2}+y^{2}}$ symmetry (see Appendix A). This pairing is thus trivial and does not support SABSs on any surface. The intraorbital component is $\underline{\Delta}_{8}(\mathbf{k})=\frac{\Delta_{0}}{A}\sigma_{0}\otimes[c_{y}(\mathbf{k})s_{1}-c_{x}(\mathbf{k})s_{2}]is_{2}$. The low energy effective model for $\underline{\Delta}_{8}(\mathbf{k})$ describes a planar state in the $k_{x}k_{y}$ plane, and thus does not support SABSs on the $xy$ surface (see Appendix A).

Because the above four pairings exhaust the pairings that can open gap within the TSSs ($xy$ surface) yet none of them support SABSs on the $xy$ surface, any pairing that supports SABSs on the $xy$ surface can not open gap within the TSSs. Thus, in the assumed situation of well defined TSSs separated from the bulk band states, the SABSs if exist will merge with the TSSs. We would like to point out that, though all pairings supporting SABSs do not open gap within the TSSs, the TSSs are modified by the pairing through coupling with the bulk states. The detailed analysis will be presented in a later work.

\section{summary}

To summarize, we propose and illustrate how the superconducting phases of the generalized BHZ model, a description of pairings in STIs and SCs realized from semiconductors with large spin orbit interaction, can be understood easily in terms of low energy effective models, constructed following the guide of analogy between TI and one orbital TSCs. Several predictions are made following this new approach. The normal state topological order is not essential for a pairing to be topologically nontrivial. Whereas the delicate couplings of degrees of freedom in the model are of crucial importance. For TI normal phase, TSCs may host one copy or two copies of SABSs, depending on the chemical potential. The previously found SABS-TSS connection structure is shown to be a universal feature of STIs.

\begin{acknowledgments}
This work is supported by NSFC.11204035 and SRFDP.20120092120040 (L.H.), and also by NSC in Taiwan under Grant No.103-2120-M-001-009 (T.K.L). Part of the calculations was performed in the National Center for High-Performance Computing in Taiwan.
\end{acknowledgments}\index{}

\emph{Note added.$\relbar$} In terms of basis transformation \cite{basis} or quasiclassical treatments \cite{qc}, some results similar to the present work were also obtained.

\begin{appendix}

\section{Low energy effective models for pairings in the generalized BHZ model of Eq.(1) of the main text}
In the main text, we have shown explicitly the deduction and full expression of the low energy effective model for $\underline{\Delta}_{1}(\mathbf{k})$. Here, we provide the construction of the low energy effective models for other numbered pairings that appear in the main text and two additional pairings, and identify them with known pairings in one-orbital model.

\emph{$\underline{\Delta}_{2}(\mathbf{k})=\Delta_{0}\sigma_{2}\otimes s_{3}$.} For this pairing, we have $h_{a}=[m(\mathbf{k})-\mu]s_{0}\otimes\tau_{3}$, $h_{b}=-[m(\mathbf{k})+\mu]s_{0}\otimes\tau_{3}$, and $h_{ab}=h^{\dagger}_{ba}=-ic_{z}(\mathbf{k})s_{0}\otimes\tau_{3}+c_{y}(\mathbf{k})s_{1}\otimes\tau_{0} -c_{x}(\mathbf{k})s_{2}\otimes\tau_{3}-i\Delta_{0}s_{3}\otimes\tau_{1}$. Hereafter in deriving the low energy effective models, we set the quasiparticle energy $E=0$, the rationality of which is to be confirmed in Sec.III of this supplemental material. So the low energy effective model within the subspace of orbital $a$ is
\begin{eqnarray}
\tilde{h}_{a}&\simeq& h_{a}-h_{ab}h^{-1}_{b}h_{ba}  \notag \\
&=&(m-\mu+\frac{c^{2}_{x}+c^{2}_{y}+c^{2}_{z}-\Delta^{2}_{0}}{m+\mu})s_{0}\otimes\tau_{3}  \notag \\ &&+\frac{2\Delta_{0}}{m+\mu}(c_{x}s_{1}\otimes\tau_{1}+c_{z}s_{3}\otimes\tau_{1}).
\end{eqnarray}
The effective pairing is still triplet but becomes $\mathbf{k}$-dependent. As is well-known, triplet pairings in a single orbital model can be written in terms of a vector $\mathbf{d}$ as $(\mathbf{d\cdot s})is_{2}=-d_{1}s_{3}+id_{2}s_{0}+d_{3}s_{1}$.\cite{salomaa87} In this notation, the above effective pairing corresponds to $d_{1}(\mathbf{k})=-\frac{2\Delta_{0}}{m(\mathbf{k})+\mu}c_{z}(\mathbf{k})$, $d_{2}=0$, and $d_{3}(\mathbf{k})=\frac{2\Delta_{0}}{m(\mathbf{k})+\mu}c_{x}(\mathbf{k})$. Close to $\mathbf{k=0}$, we have $c_{x}(\mathbf{k})\simeq Ak_{x}$ and $c_{z}(\mathbf{k})\simeq Bk_{z}$. So, apart from the prefactor $[m(\mathbf{k})+\mu]^{-1}$ which is an even function of $\mathbf{k}$ and the anisotropy coming from $A\neq B$, the effective pairing describes the planar phase in the $k_{x}k_{z}$ plane, well-known in the study of superfluid $^3$He.\cite{salomaa87} Since the effective pairing is odd in $k_{z}$, SABSs on the $xy$ surface is expected to exist, which is confirmed in previous studies.\cite{hao11,sasaki11,yamakage12} In addition, at least a pair of gap nodes exist along the $k_{y}$ axis in the surface Brillouin zone (BZ), since the effective pairing is $k_{y}$ independent. Thus the SABSs for this pairing is highly anisotropic in contrast to the SABSs for $\underline{\Delta}_{1}(\mathbf{k})$.\cite{hao11,sasaki11,yamakage12}

\emph{$\underline{\Delta}_{3}(\mathbf{k})=i\Delta_{0}\sigma_{2}\otimes s_{0}$.} This is also an interorbital triplet pairing. It has the same $h_{a}$ and $h_{b}$ as $\underline{\Delta}_{2}(\mathbf{k})$, while the interorbital couplings are $h_{ab}=h^{\dagger}_{ba}=-ic_{z}(\mathbf{k})s_{0}\otimes\tau_{3}+c_{y}(\mathbf{k})s_{1}\otimes\tau_{0} -c_{x}(\mathbf{k})s_{2}\otimes\tau_{3}+i\Delta_{0}s_{0}\otimes\tau_{2}$. The low energy effective model within the subspace of orbital $a$ is
\begin{eqnarray}
\tilde{h}_{a}&\simeq& h_{a}-h_{ab}h^{-1}_{b}h_{ba}  \notag \\
&=&(m-\mu+\frac{c^{2}_{x}+c^{2}_{y}+c^{2}_{z}-\Delta^{2}_{0}}{m+\mu})s_{0}\otimes\tau_{3}  \notag \\ &&-\frac{2\Delta_{0}}{m+\mu}(c_{y}s_{1}\otimes\tau_{1}+c_{z}s_{0}\otimes\tau_{2}).
\end{eqnarray}
In terms of the vector $\mathbf{d}$, this effective model has a pairing corresponding to $d_{1}=0$, $d_{2}(\mathbf{k})=\frac{2\Delta_{0}}{m(\mathbf{k})+\mu}c_{z}(\mathbf{k})$, and $d_{3}(\mathbf{k})=-\frac{2\Delta_{0}}{m(\mathbf{k})+\mu}c_{y}(\mathbf{k})$. Similar to the analysis for $\underline{\Delta}_{3}(\mathbf{k})$, the present effective pairing corresponds to an anisotropic planar phase in the $k_{y}k_{z}$ plane.\cite{salomaa87}

\emph{$\underline{\Delta}_{4}(\mathbf{k})=i\Delta_{0}\sigma_{1}\otimes s_{2}$.} This is an interorbital odd parity singlet pairing. For this pairing, $h_{a}$ and $h_{b}$ are the same as those for $\underline{\Delta}_{2}(\mathbf{k})$, $h_{ab}=h^{\dagger}_{ba}=-ic_{z}(\mathbf{k})s_{0}\otimes\tau_{3}+c_{y}(\mathbf{k})s_{1}\otimes\tau_{0} -c_{x}(\mathbf{k})s_{2}\otimes\tau_{3}-\Delta_{0}s_{2}\otimes\tau_{2}$. The low energy effective model within the subspace of orbital $a$ is
\begin{eqnarray}
\tilde{h}_{a}&\simeq& h_{a}-h_{ab}h^{-1}_{b}h_{ba}  \notag \\
&=&(m-\mu+\frac{c^{2}_{x}+c^{2}_{y}+c^{2}_{z}-\Delta^{2}_{0}}{m+\mu})s_{0}\otimes\tau_{3}  \notag \\ &&+\frac{2\Delta_{0}}{m+\mu}(c_{x}s_{0}\otimes\tau_{2}-c_{y}s_{3}\otimes\tau_{1}).
\end{eqnarray}
In terms of the $\mathbf{d}$ vector, this effective model has a pairing corresponding to $d_{1}=\frac{2\Delta_{0}}{m(\mathbf{k})+\mu}c_{y}(\mathbf{k})$, $d_{2}(\mathbf{k})=-\frac{2\Delta_{0}}{m(\mathbf{k})+\mu}c_{x}(\mathbf{k})$, and $d_{3}(\mathbf{k})=0$. Comparing with the effective pairings for $\underline{\Delta}_{2}(\mathbf{k})$ and $\underline{\Delta}_{3}(\mathbf{k})$, it is clear that this effective pairing also corresponds to an planar state, this time in the $k_{x}k_{y}$ plane. Since the effective pairing do not change sign under a sign reversal of $k_{z}$, this pairing does not support SABSs on the $xy$ surface, in contrast to the former two pairings.\cite{hao11}

\emph{$\underline{\Delta}_{5}(\mathbf{k})=i\Delta_{0}\varphi_{\mathbf{k}}\sigma_{2}\otimes s_{2}$.} This is a singlet pairing proposed recently by us as a possible candidate for the pairing of Cu$_{x}$Bi$_{2}$Se$_{3}$, Bi$_{2}$Te$_{3}$ and Bi$_{2}$Se$_{3}$.\cite{hao13} For this pairing, $h_{a}$ and $h_{b}$ are the same as those for $\underline{\Delta}_{2}(\mathbf{k})$, and $h_{ab}=h^{\dagger}_{ba}=-ic_{z}(\mathbf{k})s_{0}\otimes\tau_{3}+c_{y}(\mathbf{k})s_{1}\otimes\tau_{0} -c_{x}(\mathbf{k})s_{2}\otimes\tau_{3}+i\Delta_{0}\varphi_{\mathbf{k}}s_{2}\otimes\tau_{2}$.
The low energy effective model within the subspace of orbital $a$ is
\begin{eqnarray}
\tilde{h}_{a}&\simeq& h_{a}-h_{ab}h^{-1}_{b}h_{ba}  \notag \\
&=&(m-\mu+\frac{c^{2}_{x}+c^{2}_{y}+c^{2}_{z}-\Delta^{2}_{0}\varphi^{2}_{\mathbf{k}}}{m+\mu})s_{0}\otimes\tau_{3}  \notag \\
&&-\frac{2\Delta_{0}c_{z}(\mathbf{k})\varphi_{\mathbf{k}}}{m+\mu}s_{2}\otimes\tau_{2}.
\end{eqnarray}
So, the effective pairing in the low energy one-orbital model for this novel singlet pairing is simply an anisotropic singlet pairing. Since for $\mathbf{k\sim 0}$, $\varphi_{\mathbf{k}}\sim k_{x}$ or $\sim k_{y}$, the small wave vector behavior of this pairing is identical to the $d_{xz}$ or $d_{yz}$ pairing. It is thus simple to understand why this pairing supports SABSs along $x$ or $y$ directions.

\emph{$\underline{\Delta}_{6}(\mathbf{k})=i\Delta_{0}\sigma_{0}\otimes s_{2}$.} For this intraorbital singlet pairing, we have $h_{a}=[m(\mathbf{k})-\mu]s_{0}\otimes\tau_{3}-\Delta_{0}s_{2}\otimes\tau_{2}$, $h_{b}=-[m(\mathbf{k})+\mu]s_{0}\otimes\tau_{3}-\Delta_{0}s_{2}\otimes\tau_{2}$, and $h_{ab}=h^{\dagger}_{ba}=-ic_{z}(\mathbf{k})s_{0}\otimes\tau_{3}+c_{y}(\mathbf{k})s_{1}\otimes\tau_{0} -c_{x}(\mathbf{k})s_{2}\otimes\tau_{3}$. The low energy effective model within the subspace of orbital $a$ is
\begin{eqnarray}
\tilde{h}_{a}&\simeq& h_{a}-h_{ab}h^{-1}_{b}h_{ba}  \notag \\
&=&[m-\mu+(m+\mu)\frac{c^{2}_{x}+c^{2}_{y}+c^{2}_{z}}{(m+\mu)^{2}+\Delta^{2}_{0}}]s_{0}\otimes\tau_{3}  \notag \\
&&-[1+\frac{c^{2}_{x}+c^{2}_{y}+c^{2}_{z}}{(m+\mu)^{2}+\Delta^{2}_{0}}]\Delta_{0}s_{2}\otimes\tau_{2}.
\end{eqnarray}
The effective pairing is of the same symmetry and just introduces a $\mathbf{k}$ dependent renormalization and in the meantime a slight anisotropy between the dependencies on $k_{x}k_{y}$ and the dependency on $k_{z}$ to the original intraorbital pairing amplitude. So this pairing is topological trivial.

\emph{$\underline{\Delta}_{7}(\mathbf{k})= \frac{\Delta_{0}}{A}\sigma_{1}\otimes[(c_{y}(\mathbf{k})s_{1}-c_{x}(\mathbf{k})s_{2})is_{2}]$.} This pairing is interorbital triplet, with the spin part explicitly in the form of a planar state defined in the $k_{x}k_{y}$ plane. For this pairing, $h_{a}$ and $h_{b}$ are the same as those for $\underline{\Delta}_{2}(\mathbf{k})$, and $h_{ab}=h^{\dagger}_{ba}=-ic_{z}(\mathbf{k})s_{0}\otimes\tau_{3}+c_{y}(\mathbf{k})s_{1}\otimes\tau_{0} -c_{x}(\mathbf{k})s_{2}\otimes\tau_{3}+\frac{\Delta_{0}}{A}[c_{x}(\mathbf{k})s_{0}\otimes\tau_{2} -c_{y}(\mathbf{k})s_{3}\otimes\tau_{1}]$.
The low energy effective model within the subspace of orbital $a$ is
\begin{eqnarray}
\tilde{h}_{a}&\simeq& h_{a}-h_{ab}h^{-1}_{b}h_{ba}  \notag \\
&=&[m-\mu+\frac{(1-\frac{\Delta^{2}_{0}}{A^{2}})(c^{2}_{x}+c^{2}_{y})+c^{2}_{z}}{m+\mu}]s_{0}\otimes\tau_{3}
\notag \\
&&-\frac{1}{m+\mu}\frac{2\Delta_{0}}{A}(c^{2}_{x}+c^{2}_{y})s_{2}\otimes\tau_{2}.
\end{eqnarray}
The effective pairing is an anisotropic singlet pairing with the same symmetry as a $s_{x^{2}+y^{2}}$ pairing for $\mathbf{k}\sim0$. This anisotropic $s$ wave pairing do not give SABSs on any surface.

\emph{$\underline{\Delta}_{8}(\mathbf{k})= \frac{\Delta_{0}}{A}\sigma_{0}\otimes[(c_{y}(\mathbf{k})s_{1}-c_{x}(\mathbf{k})s_{2})is_{2}]$.} This is the intraorbital version of $\underline{\Delta}_{7}(\mathbf{k})$ since they have the same $\mathbf{k}$ dependency and spin subspace structure. For this pairing, $h_{a}=[m(\mathbf{k})-\mu]s_{0}\otimes\tau_{3}+\frac{\Delta_{0}}{A}[c_{x}(\mathbf{k})s_{0}\otimes\tau_{2} -c_{y}(\mathbf{k})s_{3}\otimes\tau_{1}]$, $h_{b}=-[m(\mathbf{k})+\mu]s_{0}\otimes\tau_{3}+\frac{\Delta_{0}}{A}[c_{x}(\mathbf{k})s_{0}\otimes\tau_{2} -c_{y}(\mathbf{k})s_{3}\otimes\tau_{1}]$, and $h_{ab}=h^{\dagger}_{ba}=-ic_{z}(\mathbf{k})s_{0}\otimes\tau_{3}+c_{y}(\mathbf{k})s_{1}\otimes\tau_{0} -c_{x}(\mathbf{k})s_{2}\otimes\tau_{3}$. The low energy effective model within the subspace of orbital $a$ is
\begin{eqnarray}
\tilde{h}_{a}&\simeq& h_{a}-h_{ab}h^{-1}_{b}h_{ba}  \notag \\
&=&[m-\mu+\frac{(m+\mu)(c^{2}_{x}+c^{2}_{y}+c^{2}_{z})}{(m+\mu)^{2}+\frac{\Delta^{2}_{0}}{A^{2}}(c^{2}_{x}+c^{2}_{y})}]s_{0}\otimes\tau_{3} \notag  \\
&&+\frac{\Delta_{0}}{A}[1+\frac{c^{2}_{x}+c^{2}_{y}+c^{2}_{z}}{(m+\mu)^{2}+\frac{\Delta^{2}_{0}}{A^{2}}(c^{2}_{x}+c^{2}_{y})}] \notag  \\
&&\times(c_{x}s_{0}\otimes\tau_{2}-c_{y}s_{3}\otimes\tau_{1}).
\end{eqnarray}
The correction from the mixing with the $b$ orbital just brings in a momentum dependent enhancement to the pairing within a single orbital. It is interesting to note that the effective pairing of this pairing is essentially identical with that of $\underline{\Delta}_{4}(\mathbf{k})$. This pairing thus do not support SABSs on the $xy$ surface.

The only pairing studied in the main text that is both fully gapped and supports SABSs is $\underline{\Delta}_{1}(\mathbf{k})$. The effective pairing in its low energy effective model is equivalent to the BW phase of $^{3}$He.\cite{salomaa87} It is then interesting to ask what would be the effective pairing in the low energy effective model if a BW pairing is realized in the original two-orbital model. In analogy with $\underline{\Delta}_{7}(\mathbf{k})$ and $\underline{\Delta}_{8}(\mathbf{k})$, we study one intraorbital and one interorbital BW pairing in the generalized BHZ model.

The isotropic intraorbital BW pairing could be taken as $\underline{\Delta}_{9}(\mathbf{k})=\Delta_{0}\sigma_{0}\otimes[\mathbf{d(\mathbf{k})\cdot s}]is_{2}$, in which $d_{1}(\mathbf{k})=c_{x}(\mathbf{k})/A$, $d_{2}(\mathbf{k})=c_{y}(\mathbf{k})/A$, and $d_{3}(\mathbf{k})=c_{z}(\mathbf{k})/B$. For this pairing, $h_{a}=[m(\mathbf{k})-\mu]s_{0}\otimes\tau_{3}-\Delta_{0}[d_{1}(\mathbf{k})s_{3}\otimes\tau_{1} +d_{2}(\mathbf{k})s_{0}\otimes\tau_{2}-d_{3}(\mathbf{k})s_{1}\otimes\tau_{1}]$, $h_{b}=-[m(\mathbf{k})+\mu]s_{0}\otimes\tau_{3}-\Delta_{0}[d_{1}(\mathbf{k})s_{3}\otimes\tau_{1} +d_{2}(\mathbf{k})s_{0}\otimes\tau_{2}-d_{3}(\mathbf{k})s_{1}\otimes\tau_{1}]$, and $h_{ab}=h^{\dagger}_{ba}=-ic_{z}(\mathbf{k})s_{0}\otimes\tau_{3}+c_{y}(\mathbf{k})s_{1}\otimes\tau_{0} -c_{x}(\mathbf{k})s_{2}\otimes\tau_{3}$. The low energy effective model within the subspace of orbital $a$ is
\begin{eqnarray}
\tilde{h}_{a}&\simeq& h_{a}-h_{ab}h^{-1}_{b}h_{ba}  \notag \\
&=&h_{a}+\frac{(m+\mu)(c^{2}_{x}+c^{2}_{y}+c^{2}_{z})}{(m+\mu)^{2} +\Delta^{2}_{0}(d^{2}_{1}+d^{2}_{2}+d^{2}_{3})}s_{0}\otimes\tau_{3}  \notag \\
&&-\frac{(2A+B)c^{2}_{z}-B(c^{2}_{x}+c^{2}_{y})}{(m+\mu)^{2} +\Delta^{2}_{0}(d^{2}_{1}+d^{2}_{2}+d^{2}_{3})}\frac{\Delta_{0}}{AB}  \notag \\ &&\times(c_{x}s_{3}\otimes\tau_{1}+c_{y}s_{0}\otimes\tau_{2})  \\
&&+\frac{Ac^{2}_{z}-(A+2B)(c^{2}_{x}+c^{2}_{y})}{(m+\mu)^{2} +\Delta^{2}_{0}(d^{2}_{1}+d^{2}_{2}+d^{2}_{3})}\frac{\Delta_{0}}{AB}c_{z}s_{1}\otimes\tau_{1}.  \notag
\end{eqnarray}
It is clear that, besides the BW pairing inherited directly from $h_{a}$, the mixing with $b$ orbital only introduces some anisotropy between the $k_{x}k_{y}$ plane and the $k_{z}$ direction into the otherwise isotropic pairing.

The isotropic interorbital BW pairing is taken as $\underline{\Delta}_{10}(\mathbf{k})=\Delta_{0}\sigma_{1}\otimes[\mathbf{d(\mathbf{k})\cdot s}]is_{2}$. The $\mathbf{d}$ vector is taken as identical to that of $\underline{\Delta}_{9}(\mathbf{k})$. For this pairing, $h_{a}=[m(\mathbf{k})-\mu]s_{0}\otimes\tau_{3}$, $h_{b}=-[m(\mathbf{k})+\mu]s_{0}\otimes\tau_{3}$, and $h_{ab}=h^{\dagger}_{ba}=-ic_{z}(\mathbf{k})s_{0}\otimes\tau_{3}+c_{y}(\mathbf{k})s_{1}\otimes\tau_{0} -c_{x}(\mathbf{k})s_{2}\otimes\tau_{3}-\Delta_{0}[d_{1}(\mathbf{k})s_{3}\otimes\tau_{1} +d_{2}(\mathbf{k})s_{0}\otimes\tau_{2}-d_{3}(\mathbf{k})s_{1}\otimes\tau_{1}]$. The low energy effective model within the subspace of orbital $a$ is
\begin{eqnarray}
\tilde{h}_{a}&\simeq& h_{a}-h_{ab}h^{-1}_{b}h_{ba}  \notag \\
&=&[m-\mu+\frac{c^{2}_{x}+c^{2}_{y}+c^{2}_{z}-\Delta^{2}_{0}(d^{2}_{1}+d^{2}_{2}+d^{2}_{3})}{m+\mu}]s_{0}\otimes\tau_{3}
\notag \\
&&+\frac{2\Delta_{0}}{m+\mu}(c_{x}d_{y}-c_{y}d_{x})s_{2}\otimes\tau_{2}.
\end{eqnarray}
For a generally chosen set of $d_{\alpha}(\mathbf{k})\sim k_{\alpha}$ ($\alpha=x,y,z$), the effective pairing is singlet with $d_{xy}$ symmetry. However, for our above ansatz for $d_{\alpha}(\mathbf{k})$, the effective pairing in fact vanishes. Thus, the low energy effective model is a one-orbital model without of pairing. The original pairing only makes slight modifications to the effective band structure.

\section{Low energy effective models for pairings in the modified Eq.(1)}
We have used the low energy effective pairings for two typical pairings realized in the modified Eq.(1) to illustrate the importance of intricate coupling among the degrees of freedom in the model in determining the low energy effective model. The modification is introduced by replacing the term $c_{z}(\mathbf{k})\sigma_{2}\otimes s_{0}$ in Eq.(1) of the main text by $c_{z}(\mathbf{k})\sigma_{1}\otimes s_{3}$.

The first pairing is $\underline{\Delta}_{1}(\mathbf{k})=\Delta_{0}\sigma_{2}\otimes s_{1}$. For the model defined as modified Eq.(1), we have $h_{a}=[m(\mathbf{k})-\mu]s_{0}\otimes\tau_{3}$, $h_{b}=-[m(\mathbf{k})+\mu]s_{0}\otimes\tau_{3}$, and $h_{ab}=h^{\dagger}_{ba}=c_{z}(\mathbf{k})s_{3}\otimes\tau_{0}+c_{y}(\mathbf{k})s_{1}\otimes\tau_{0} -c_{x}(\mathbf{k})s_{2}\otimes\tau_{3}-i\Delta_{0}s_{1}\otimes\tau_{1}$. The low energy effective model within the subspace of orbital $a$ is
\begin{eqnarray}
\tilde{h}_{a}&\simeq& h_{a}-h_{ab}h^{-1}_{b}h_{ba}  \notag \\
&=&[m-\mu+\frac{c^{2}_{x}+c^{2}_{y}+c^{2}_{z}-\Delta^{2}_{0}}{m+\mu}]s_{0}\otimes\tau_{3}  \notag \\
&&-\frac{2\Delta_{0}}{m+\mu}[c_{x}s_{3}\otimes\tau_{1}+c_{y}s_{0}\otimes\tau_{2}].
\end{eqnarray}
In terms of the $\mathbf{d}$ vector, the effective pairing is $d_{1}(\mathbf{k})=\frac{2\Delta_{0}}{m(\mathbf{k})+\mu}c_{x}(\mathbf{k})$, $d_{2}(\mathbf{k})=\frac{2\Delta_{0}}{m(\mathbf{k})+\mu}c_{y}(\mathbf{k})$, and $d_{3}=0$. This pairing is thus equivalent to the planar phase of $^{3}$He.\cite{salomaa87} However, we know in the main text that, the low energy effective pairing for this pairing is equivalent to the BW phase of $^{3}$He for the original model of Eq.(1).

The other pairing that we study in the main text for the modified Eq.(1) is $\underline{\Delta}_{4}(\mathbf{k})=i\Delta_{0}\sigma_{1}\otimes s_{2}$. For this pairing we have $h_{a}=[m(\mathbf{k})-\mu]s_{0}\otimes\tau_{3}$, $h_{b}=-[m(\mathbf{k})+\mu]s_{0}\otimes\tau_{3}$, and $h_{ab}=h^{\dagger}_{ba}=c_{z}(\mathbf{k})s_{3}\otimes\tau_{0}+c_{y}(\mathbf{k})s_{1}\otimes\tau_{0} -c_{x}(\mathbf{k})s_{2}\otimes\tau_{3}-\Delta_{0}s_{2}\otimes\tau_{2}$. The low energy effective model within the subspace of orbital $a$ is
\begin{eqnarray}
\tilde{h}_{a}&\simeq& h_{a}-h_{ab}h^{-1}_{b}h_{ba}  \notag \\
&=&[m-\mu+\frac{c^{2}_{x}+c^{2}_{y}+c^{2}_{z}-\Delta^{2}_{0}}{m+\mu}]s_{0}\otimes\tau_{3}  \notag \\
&&+\frac{2\Delta_{0}}{m+\mu}[c_{x}s_{0}\otimes\tau_{2}-c_{y}s_{3}\otimes\tau_{1}+c_{z}s_{1}\otimes\tau_{1}].
\end{eqnarray}
In terms of the $\mathbf{d}$ vector, the effective triplet pairing is characterized by $d_{1}(\mathbf{k})=\frac{2\Delta_{0}}{m(\mathbf{k})+\mu}c_{y}(\mathbf{k})$, $d_{2}(\mathbf{k})=-\frac{2\Delta_{0}}{m(\mathbf{k})+\mu}c_{x}(\mathbf{k})$, and $d_{3}(\mathbf{k})=\frac{2\Delta_{0}}{m(\mathbf{k})+\mu}c_{z}(\mathbf{k})$. Compare with Eq.(3) of the main text and the related discussion, it is clear that this effective pairing is equivalent to the BW phase of $^{3}$He.\cite{salomaa87}

Summing up the above results and compare them with those for the original model of Eq.(1), it is clear that a slight change of the model respecting the topological nature of the normal phase can bring drastic changes to the effective pairing symmetry and the SABSs of a certain pairing expressed in the full two-orbital basis.

\section{Corrections to the low energy effective model from the quasiparticle energy $E$}
In obtaining the low energy effective models within the $a$ orbital subspace, we have neglected the quasiparticle energy $E$ in $(E-h_{b})^{-1}$ of the full formula $\tilde{h}_{a}=h_{a}+h_{ab}(E-h_{b})^{-1}h_{ba}$. We now show that, since we are concerned only with low energy excitations, the neglected terms are indeed small quantities for small pairing amplitudes.

Take $\underline{\Delta}_{1}(\mathbf{k})$ as an example, we have $h_{a}=[m(\mathbf{k})-\mu]s_{0}\otimes\tau_{3}$, $h_{b}=-[m(\mathbf{k})+\mu]s_{0}\otimes\tau_{3}$, and $h_{ab}=h^{\dagger}_{ba}=-ic_{z}(\mathbf{k})s_{0}\otimes\tau_{3}+c_{y}(\mathbf{k})s_{1}\otimes\tau_{0} -c_{x}(\mathbf{k})s_{2}\otimes\tau_{3}-i\Delta_{0}s_{1}\otimes\tau_{1}$. The term relevant to $E$ is
\begin{equation}
(E-h_{b})^{-1}=\frac{1}{m+\mu}
\begin{pmatrix} \frac{1}{1+\frac{E}{m+\mu}} & 0 & 0 & 0  \\
0 & \frac{1}{\frac{E}{m+\mu}-1} & 0 & 0 \\
0 & 0 & \frac{1}{1+\frac{E}{m+\mu}} & 0 \\
0 & 0 & 0 & \frac{1}{\frac{E}{m+\mu}-1} \end{pmatrix}.
\end{equation}
The cases that we focus in this work are characterized by $\mu\gg\Delta_{0}>0$ and $m\approx\mu$ for the wave vectors relevant to pairing. So we have $m+\mu\gg\Delta_{0}$. Since we focus on low energy excitations, in particular within the gap, the quasiparticle energy $E$ is at most the same order as $\Delta_{0}$. So the condition $|\frac{E}{m+\mu}|\ll1$ holds. Thus it is reasonable to set $E=0$ directly in Eq.(1) of the main text when analyzing low energy properties.

We can expand the diagonal elements of $(E-h_{b})^{-1}$ into Taylor series of $\frac{E}{m+\mu}$ to see more clearly the correction by $E$. It turns out that the even and odd polynomials of $\frac{E}{m+\mu}$ form two qualitatively different contributions. They are rearranged into two infinite series which after resummation yield
\begin{eqnarray}
&&(E-h_{b})^{-1}=-\frac{E}{(m+\mu)^{2}}\frac{1}{1-\frac{E^{2}}{(m+\mu)^{2}}} s_{0}\otimes\tau_{0} \notag \\
&&+\frac{1}{m+\mu}[1+\frac{E^{2}}{(m+\mu)^{2}}\frac{1}{1-\frac{E^{2}}{(m+\mu)^{2}}}] s_{0}\otimes\tau_{3}.
\end{eqnarray}
After substituting into $\tilde{h}_{a}$, the odd order corrections, which are dominated by the first order term, gives a band shift of the order of $E$ to all the four Nambu orbitals within the $a$ subspace. Except for very critical parameters which we disregard in the present work, this tiny band shift does not influence the PHI of the effective model, and is thus topologically trivial. On the other hand, the correction of $\frac{E}{m+\mu}$ to the topologically relevant $s_{0}\otimes\tau_{3}$ term comes in the second order and thus is even more safely negligible for $|\frac{E}{m+\mu}|\ll1$.

\section{Influence of the band shift term}

The $\epsilon(\mathbf{k})\sigma_{0}\otimes s_{0}$ term in Eq.(1) of the main text represents a band shift irrelevant to the topological nature of the normal phase. To make the analogy between TIs and one-orbital TSCs more exact, we have neglected this term in the deductions and discussions. Here, taking $\underline{\Delta}_{1}(\mathbf{k})=\Delta_{0}\sigma_{2}\otimes s_{1}$ as an example, we see if this term can bring any significant changes to our former conclusion in the main text and in this supplemental material.

Since $\epsilon(\mathbf{k})$ is even in $\mathbf{k}$, the presence of the band shift term amounts to replacing the chemical potential $\mu$ in the BdG Hamiltonian by $\mu-\epsilon(\mathbf{k})$. Thus, the corresponding low energy effective model is easily obtained by replacing $\mu$ in the original effective model in Eq.(3) of the main text by $\mu-\epsilon(\mathbf{k})$, which is
\begin{eqnarray}
&&\tilde{h}_{a}\simeq
[m-\mu+\epsilon+\frac{c^{2}_{x}+c^{2}_{y}+c^{2}_{z} -\Delta^{2}_{0}}{m+\mu-\epsilon}]s_{0}\otimes\tau_{3}   \notag \\
&&-\frac{2\Delta_{0}}{m+\mu-\epsilon} [c_{x}s_{3}\otimes\tau_{1}+c_{y}s_{0}\otimes\tau_{2}-c_{z}s_{1}\otimes\tau_{1}].
\end{eqnarray}
Following the main text, we have made the wave vector dpendencies of the various terms in the model implicit. The form of the effective pairing is obviously unchanged. So, if $\epsilon(\mathbf{k})$ is of any significance, the impact should manifest through the band energy term and appear as an influence on the PHI of the effective model.

Before analyzing the effect of $\epsilon(\mathbf{k})$ on PHI, it is beneficial to mention why for $\epsilon(\mathbf{k})=0$ and small $\Delta_{0}$ the PHI always occurs for the low energy effective model, once $m-\mu$ could change sign (i.e., PHI occurs for orbital $a$). From the definition of $m(\mathbf{k})$ in the main text, we know $m\geq m_{0}$ and $m(\mathbf{k}=\mathbf{0})=m_{0}$. We have taken $\mu>|m_{0}|$ in most of time to make sure that the PHI occurs only in the $a$ orbital. Then, for $\mathbf{k}=\mathbf{0}$ we have $\tilde{m}(\mathbf{k}=\mathbf{0})=m-\mu+\frac{c^{2}_{x}+c^{2}_{y}+c^{2}_{z} -\Delta^{2}_{0}}{m+\mu}=m_{0}-\mu-\frac{\Delta^{2}_{0}}{m_{0}+\mu}<m_{0}-\mu<0$. For $m-\mu=0$ we have $\tilde{m}(\mathbf{k})=\frac{c^{2}_{x}+c^{2}_{y}+c^{2}_{z}-\Delta^{2}_{0}}{2\mu}|_{m(\mathbf{k})=\mu}$, which is already larger than zero once $\Delta_{0}$ is small enough.

Now turn on $\epsilon(\mathbf{k})$, the question becomes: If $m(\mathbf{k})+\epsilon(\mathbf{k})-\mu$ changes sign somewhere in the BZ around $\boldsymbol{\Gamma}$, whether or not $\tilde{m}^{\prime}(\mathbf{k})=\tilde{m}(\mathbf{k})|_{\mu\rightarrow\mu-\epsilon(\mathbf{k})}=m+\epsilon-\mu +\frac{c^{2}_{x}+c^{2}_{y}+c^{2}_{z}-\Delta^{2}_{0}}{m-\epsilon+\mu}$ still changes sign somewhere in the BZ around the $\boldsymbol{\Gamma}$ point. We set $\epsilon_{0}<0$, $\epsilon_{1}>0$ and $\epsilon_{2}>0$, in agreement with the parameters obtained by fitting first principle band structures.\cite{liu10,hao13} If $\epsilon_{\alpha}$ are all much smaller than $m_{\alpha}$, $\epsilon(\mathbf{k})$ can not change the above conclusion obtained for $\epsilon(\mathbf{k})=0$, except for critical parameters that we disregard for which $m-\mu$ is on the verge of no sign change. However, in the opposite case, when the magnitudes of $\epsilon_{\alpha}$ are all much larger than $m_{\alpha}$, the sign change in $m+\epsilon-\mu$ would mostly because of the sign change of $\epsilon-\mu$. For $\mathbf{k}=\mathbf{0}$, we still have $\tilde{m}^{\prime}(\mathbf{k}=\mathbf{0})=m_{0}+\epsilon_{0}-\mu-\frac{\Delta^{2}_{0}}{m_{0}-\epsilon_{0}+\mu}<0$, since we have assumed $\mu>m_{0}+\epsilon_{0}$, $\mu>0$ and $\epsilon_{0}<m_{0}<0$. Now consider the portion of the BZ where sign change of $m+\epsilon-\mu$ occurs. In the present case of $\epsilon(\mathbf{k})$ dominating over $m(\mathbf{k})$, we have $m(\mathbf{k})\approx m_{0}$ for $m+\epsilon-\mu\approx0$, which gives $\mu-\epsilon\approx m_{0}$. Thus for $m+\epsilon-\mu\approx0$, we now have $\tilde{m}^{\prime}(\mathbf{k})\approx\frac{c^{2}_{x}+c^{2}_{y}+c^{2}_{z} -\Delta^{2}_{0}}{2m_{0}}|_{m_{0}+\epsilon-\mu\approx0}$. For very small $\Delta_{0}$ the above formula is smaller than zero. So, in comparison to the case of $\epsilon=0$, the PHI in the low energy effective model becomes harder to occur in the presence of $\epsilon(\mathbf{k})$ term. Thus, the band shift represented by $\epsilon(\mathbf{k})$ turn to destroy the TSC through forbidding the occurrence of PHI in the low energy effective model. But since in actual materials, the $\epsilon(\mathbf{k})$ term is at most of the same order of magnitude as the $m(\mathbf{k})$ term, there is still a sufficient wide range of $\mu$ which can make the superconducting state topologically nontrivial if $\underline{\Delta}_{1}(\mathbf{k})$ is realized in the full model.\cite{liu10,hao13}

\section{Conditions for the occurrence of PHI in both of two orbitals}

Though we discuss in this work mostly in terms of the $a$ and $b$ orbitals with separately even and odd parities, we also require the chemical potential $\mu$ to cross the bulk energy bands since we are interested in bulk pairing. If we neglect the band shift term in Eq.(1) of the main text, $\epsilon(\mathbf{k})\sigma_{0}\otimes s_{0}$, the dispersion of the bulk conduction band is\cite{hao11}
\begin{equation}
E(\mathbf{k})=\sqrt{m^{2}(\mathbf{k})+c_{x}^{2}(\mathbf{k})+c_{y}^{2}(\mathbf{k})+c_{z}^{2}(\mathbf{k})}.
\end{equation}
Close to the BZ center $\mathbf{k}\simeq\mathbf{0}$, we can approximate the terms in $E(\mathbf{k})$ as $m(\mathbf{k})\simeq m_{0}+\frac{3}{2}m_{1}(k^{2}_{x}+k^{2}_{y})+m_{2}k^{2}_{z}$, $c_{x}(\mathbf{k})\simeq Ak_{x}$, $c_{y}(\mathbf{k})\simeq Ak_{y}$, and $c_{z}(\mathbf{k})\simeq Bk_{z}$. For a TI, we have $m_{0}m_{1}<0$ and $m_{1}m_{2}>0$.

Since the overlap region of $m(\mathbf{k})$ ($a$ orbital) and $-m(\mathbf{k})$ ($b$ orbital) is specified by $|m(\mathbf{k})|<|m_{0}|$, the chemical potential $\mu$ must be tuned to this region to ensure that PHI occurs for both of the two orbitals. In order to make this realistic, we demand that $E(\mathbf{k})<|m_{0}|$ must be true in some part of the BZ. Expanding $E^{2}(\mathbf{k})$ in the neighborhood of $\mathbf{k}=\mathbf{0}$ into polynomials of $k_{\alpha}$ ($\alpha$ is $x$, $y$ or $z$), it is easy to see that the above condition is met if and only if at least one of the following two conditions are satisfied: $A^{2}+3m_{0}m_{1}<0$ and $B^{2}+2m_{0}m_{2}<0$. The two conditions are equivalently written as $A^{2}<3|m_{0}m_{1}|$ and $B^{2}<2|m_{0}m_{2}|$.

If $A^{2}<3|m_{0}m_{1}|$ \emph{and} $B^{2}>2|m_{0}m_{2}|$, minimum of the bulk conduction band occurs at $\sqrt{k^{2}_{x}+k^{2}_{y}}>0$ and $k_{z}=0$. If $A^{2}>3|m_{0}m_{1}|$ \emph{and} $B^{2}<2|m_{0}m_{2}|$, minimum of the bulk conduction band occurs at $k_{x}=k_{y}=0$ and $k_{z}\ne0$. Whereas for $A^{2}>3|m_{0}m_{1}|$ \emph{and} $B^{2}>2|m_{0}m_{2}|$, minimum of the bulk conduction band occurs at the BZ center $k_{x}=k_{y}=k_{z}=0$ and is equal to $|m_{0}|$, so the PHI can occur only for one orbital. In Figs.2(c) and 2(d) of the main text, we have shown surface spectral functions for two parameter sets satisfying $A^{2}>3|m_{0}m_{1}|$ \emph{and} $B^{2}<2|m_{0}m_{2}|$. Fig.1(c) of the main text is a schematic illustration of the relative positions of $\pm m(\mathbf{k})$ and the bulk conduction and valence bands, for parameter sets similar to Figs.2(c) and 2(d).

\section{More discussions on cases with two pairs of SABSs}

For parameters similar to Figs.2(a) and 2(b) of the main text (band structures of which are as illustrated in Figs.1(a) and 1(b) of the main text), PHI occurs only for one orbital and a low energy effective model related mainly to that orbital is obtained, as is shown in Eq.(3) of the main text and the first and second sections of this supplemental material. From the low energy effective model, we can tell whether the pairing is topologically nontrivial and supports SABSs. The SABSs, if they exist, can then be constructed in the same way as the topological surface states are constructed for a topological insulator.\cite{hao11,liu10}

For parameters similar to Figs.2(c) and 2(d) of the main text (the schematic illustration for which are shown in Fig.1(c) of the main text), PHI occurs for both of the two orbitals. We thus have two low energy effective models, $\tilde{h}_{a}$ related mainly to orbital $a$ and $\tilde{h}_{b}$ related mainly to orbital $b$, which are defined separately in BZ$_{a}$ and BZ$_{b}$. Suppose both of the two effective models (i.e., $\tilde{h}_{a}$ and $\tilde{h}_{b}$) can give rise to SABSs on the $xy$ plane, such as $\underline{\Delta}_1(\mathbf{k})$ and $\underline{\Delta}_2(\mathbf{k})$ realized in Eq.(1) of the main text.

Upon the introduction of a surface perpendicular to the $z$ direction, while on one hand we expect that $\tilde{h}_{a}$ and $\tilde{h}_{b}$ will each yield a pair of SABSs, on the other hand states in BZ$_{a}$ and states in BZ$_{b}$ (and thus the two pairs of SABSs emerging separately from $\tilde{h}_{a}$ and $\tilde{h}_{b}$) will be coupled together through the mixing of $k_{z}$, since the translational invariance along $z$ is broken. This is the general picture to understand the final dispersion of the SABSs shown in Figs.2(c) and 2(d). Correspondingly, we take two steps to get the SABSs in Figs.2(c) and 2(d) by hand: Firstly, calculate the SABSs for $\tilde{h}_{a}$ and $\tilde{h}_{b}$ separately by ignoring their coupling. Secondly, estimate the coupling between the two pairs of SABSs and thus get the final dispersion for the true SABSs.

In what follows, we first make a general analysis of the physical ingredients relevant to the formation of SABSs in Figs.2(c) and 2(d). Based on this analysis, we then make a rough estimation of the coupling strength between $\tilde{h}_{a}$ and $\tilde{h}_{b}$ as a result of broken translational invariance induced by the surface, which is compared to the numerical surface spectrum. Finally, we show explicitly how the SABSs can be obtained analytically. In these analyses, we always assume that the pairing amplitude is much smaller than other parameters, such as the chemical potential.

\emph{\textbf{- A general analysis.}} We have shown in the above section that, when $A^{2}<3|m_{0}m_{1}|$ or $B^{2}<2|m_{0}m_{2}|$, PHI can occur for both orbital $a$ and orbital $b$ when $|\mu/m_{0}|<1$. For realistic $\mu$ crossing the bulk band and small pairing amplitude $\Delta_{0}$ (i.e., $|\Delta_{0}/\mu|\ll1$), we can get two low energy effective models $\tilde{h}_{a}$ and $\tilde{h}_{b}$, defined within BZ$_{a}$ and BZ$_{b}$. The expression for $\tilde{h}_{a}$ can be found from Eq.(3) of the main text, or from Sec.I and Sec.II of this supplemental material, for different pairings. $\tilde{h}_{b}$ can be obtained by making proper substitutions to $\tilde{h}_{a}$. For example, for the first pairing formed in Eq.(1) of the main text, we have $\tilde{h}_{b}=\tilde{h}_{a}[m\rightarrow-m,\Delta_{0}\rightarrow-\Delta_{0},c_{z}\rightarrow-c_{z}]$, whereas we have $\tilde{h}_{b}=\tilde{h}_{a}[m\rightarrow-m,\Delta_{0}\rightarrow-\Delta_{0}]$ for the first pairing formed in the modified Eq.(1) of the main text. BZ$_{a}$ consists of states of orbital $a$ ensuring $|m(\mathbf{k})-\mu|$ to be within the order of $\Delta_{0}$, while BZ$_{b}$ consists of states of orbital $b$ ensuring $|-m(\mathbf{k})-\mu|$ to be within the order of $\Delta_{0}$.

To facilitate the following analysis, we reexpress the wave vectors in the BZ as $\mathbf{k}=(\mathbf{k}_{xy},k_{z})$, where $\mathbf{k}_{xy}=(k_{x},k_{y})$. For each fixed $\mathbf{k}_{xy}$, $m(\mathbf{k}_{xy},k_{z})$ is an even function of $k_{z}$. So, we can further separate BZ$_{a}$ (and also BZ$_{b}$) into two portions with positive and negative $k_{z}$, respectively. They are denoted as BZ$^{+}_{a}$ (BZ$^{+}_{b}$) and BZ$^{-}_{a}$ (BZ$^{-}_{b}$). Considering BZ$_{a}$, for a certain $\mathbf{k}_{xy}$, define $k_{z0}\ge0$ as the solution of $m(\mathbf{k}_{xy},k_{z0})-\mu=0$. Then BZ$^{+}_{a}$ contains a slice of states with the same $\mathbf{k}_{xy}$, centering around $(\mathbf{k}_{xy},k_{z0})$ and having approximately a number of
\begin{equation}
\frac{\nu_{a}\Delta_{0}/|v_{z}(\mathbf{k}_{xy},k_{z0})|}{2\pi/N_{z}}=\frac{\nu_{a}N_{z}\Delta_{0}}{2\pi |v_{z}(\mathbf{k}_{xy},k_{z0})|}=N^{a+}_{z}(\mathbf{k}_{xy})
\end{equation}
states, where $N_{z}$ is the number of quintuple layers along the $z$ direction and $v_{z}(\mathbf{k}_{xy},k_{z0})=\frac{\partial}{\partial k_{z}}m(\mathbf{k})|_{\mathbf{k}=(\mathbf{k}_{xy},k_{z0})}$. $\nu_{a}$ is a number of order 1, indicating the effective acting range of the pairing correlation in BZ$^{+}_{a}$. BZ$^{-}_{a}$ contains a corresponding slice of states centered at $(\mathbf{k}_{xy},-k_{z0})$ and having the same number of states. Similarly, BZ$^{+}_{b}$ contains for the same $\mathbf{k}_{xy}$ a slice of states, centering around $(\mathbf{k}_{xy},k^{'}_{z0})$ and having approximately a number of
\begin{equation}
\frac{\nu_{b}\Delta_{0}/|v_{z}(\mathbf{k}_{xy},k^{'}_{z0})|}{2\pi/N_{z}}=\frac{\nu_{b}N_{z}\Delta_{0}}{2\pi |v_{z}(\mathbf{k}_{xy},k^{'}_{z0})|}=N^{b+}_{z}(\mathbf{k}_{xy})
\end{equation}
states, where $k^{'}_{z0}\ge0$ is solution of $m(\mathbf{k}_{xy},k^{'}_{z0})+\mu=0$ and $v_{z}(\mathbf{k}_{xy},k^{'}_{z0})=\frac{\partial}{\partial k_{z}}m(\mathbf{k})|_{\mathbf{k}=(\mathbf{k}_{xy},k^{'}_{z0})}$.  $\nu_{b}$ is a number of order 1, indicating the effective acting range of the pairing correlation in BZ$^{+}_{b}$. BZ$^{-}_{b}$ contains a corresponding slice of states centered at $(\mathbf{k}_{xy},-k^{'}_{z0})$ and having the same number of states as BZ$^{+}_{b}$. If for a certain $\mathbf{k}_{xy}$, we have $k_{z0}\simeq0$ ($k^{'}_{z0}\simeq0$), then for this $\mathbf{k}_{xy}$ BZ$^{+}_{a}$ and B$Z^{-}_{a}$ (BZ$^{+}_{b}$ and BZ$^{-}_{b}$) each has half the number of states as specified above.

For an ideal bulk material with periodic boundary conditions in all three directions, states within BZ$_{a}$ and BZ$_{b}$ are independent of each other. The introduction of two $xy$ surfaces break the translational invariance along $z$ direction, so that $k_{z}$ is not still a good quantum number and states in BZ$_{a}$ and BZ$_{b}$ will be connected through the mixing of $k_{z}$. Since we are here only concerned with the coupling between orbital $a$ within BZ$_{a}$ and orbital $b$ within BZ$_{b}$, we illustrate the effect of $k_{z}$ mixing by focusing on the terms in the BdG Hamiltonian that couple orbitals $a$ and $b$ along the $z$ direction, which are
\begin{equation}
h^{z}_{ab}=-i\sum\limits_{\mathbf{k}}\phi^{\dagger}_{\mathbf{k}a}c_{z}(\mathbf{k})s_{0}\otimes\tau_{3} \phi_{\mathbf{k}b}+h.c.,
\end{equation}
for the bulk material. Turning the $z$ direction of $h^{z}_{ab}$ to the real space by making the Fourier transformation (the lattice parameters have been set as length units)
\begin{equation}
\phi_{\mathbf{k}a(b)}=\frac{1}{\sqrt{N_{z}}}\sum\limits^{N_{z}}_{n_{z}=1}\phi_{\mathbf{k}_{xy}n_{z}a(b)}e^{-ik_{z}n_{z}},
\end{equation}
we get
\begin{equation}
h^{z}_{ab}=\frac{B}{2}\sum\limits_{\mathbf{k}_{xy}}\sum\limits^{N_{z}}_{n_{z}=1}\sum\limits_{\alpha=\pm1}
\phi^{\dagger}_{\mathbf{k}_{xy}n_{z}a}\alpha s_{0}\otimes\tau_{3} \phi_{\mathbf{k}_{xy},n_{z}-\alpha,b}+h.c.
\end{equation}
Until now, we are considering a fully three dimensional bulk material. The periodic boundary condition along $z$ direction is enforced by requiring that $n_{z}=0$ is equivalent to $n_{z}=N_{z}$ and $n_{z}=N_{z}+1$ is equivalent to $n_{z}=1$.

A sample with two $xy$ surfaces are obtained by cutting off the coupling between the $n_{z}=1$ and the $n_{z}=N_{z}$ quintuple layers. $h^{z}_{ab}$ then becomes
\begin{eqnarray}
\bar{h}^{z}_{ab}&=&\frac{B}{2}\sum\limits_{\mathbf{k}_{xy}}\sum\limits^{N_{z}-1}_{n_{z}=1}
[\phi^{\dagger}_{\mathbf{k}_{xy},n_{z}+1,a} s_{0}\otimes\tau_{3}\phi_{\mathbf{k}_{xy}n_{z}b}  \notag \\
&&-\phi^{\dagger}_{\mathbf{k}_{xy}n_{z}a} s_{0}\otimes\tau_{3}\phi_{\mathbf{k}_{xy},n_{z}+1,b}]+h.c.
\end{eqnarray}
The mixing of $k_{z}$ quantum states defined in the original bulk model is explicitly constructed by making the inverse Fourier transformation of Eq.(F4) to $\bar{h}^{z}_{ab}$, which yields after completing the summation over $n_{z}$
\begin{equation}
\bar{h}^{z}_{ab}=\sum\limits_{\mathbf{k}_{xy}}\frac{iB}{N_{z}}\sum\limits_{k_{z}k^{'}_{z}}\phi^{\dagger}_{\mathbf{k}_{xy}k_{z}a} f_{k_{z}k^{'}_{z}} s_{0}\otimes\tau_{3}\phi_{\mathbf{k}_{xy}k^{'}_{z}b}+h.c.,
\end{equation}
where
\begin{equation}
f_{k_{z}k^{'}_{z}}=\frac{1}{2i}[e^{ik^{'}_{z}}-e^{-ik_{z}}]=\sin\frac{k_{z}+k^{'}_{z}}{2}e^{-i\frac{k_{z}-k^{'}_{z}}{2}}.
\end{equation}
Note that, if we are considering a sample running from $n_{z}=-N_{z}$ to $n_{z}=-1$, the same procedure as done above shows that Eq.(F7) keeps invariant, whereas Eq.(F8) is replaced by
\begin{equation}
f_{k_{z}k^{'}_{z}}=\frac{1}{2i}[e^{ik_{z}}-e^{-ik^{'}_{z}}]=\sin\frac{k_{z}+k^{'}_{z}}{2}e^{i\frac{k_{z}-k^{'}_{z}}{2}}.
\end{equation}
The coupling strength between states in BZ$_{a}$ and BZ$_{b}$ by the introduction of two $xy$ surfaces can thus be estimated from  Eqs.(F7) to (F9).

\emph{\textbf{- Rough estimation of average coupling strength between states in BZ$_{a}$ and BZ$_{b}$.}}
We now estimate the average coupling strength between states of orbital $a$ in BZ$^{\alpha}_{a}$ ($\alpha=\pm$) and states of orbital $b$ in BZ$^{\beta}_{b}$ ($\beta=\pm$), induced by the creation of a pair of $xy$ surfaces. Since $k_{x}$ and $k_{y}$ are still good quantum numbers after a pair of $xy$ surfaces are introduced, states pertaining to different $\mathbf{k}_{xy}$ are independent, so we estimate the effective coupling for a certain $\mathbf{k}_{xy}$ shared by BZ$_{a}$ and BZ$_{b}$. Suppose the states in BZ$^{+}_{a}$ centering around $(\mathbf{k}_{xy},k_{z0})$ and the states in BZ$^{-}_{a}$ centering around $(\mathbf{k}_{xy},-k_{z0})$ are far away from each other, and the states in BZ$^{+}_{b}$ centering around $(\mathbf{k}_{xy},k^{'}_{z0})$ and the states in BZ$^{-}_{b}$ centering around $(\mathbf{k}_{xy},-k^{'}_{z0})$ are also far away from each other. This assumption is reasonable for most $\mathbf{k}_{xy}$ (including center of the surface BZ $k_{x}=k_{y}=0$), since we assume the pairing amplitude is very small. Then the coupling between states in BZ$_{a}$ and states in BZ$_{b}$ for $\mathbf{k}_{xy}$ are
\begin{eqnarray}
&&\bar{\bar{h}}^{z}_{ab}(\mathbf{k}_{xy})=\frac{iB}{N_{z}}\sum\limits_{k_{z}k^{'}_{z}}\phi^{\dagger}_{\mathbf{k}_{xy}k_{z}a} f_{k_{z}k^{'}_{z}} s_{0}\otimes\tau_{3}\phi_{\mathbf{k}_{xy}k^{'}_{z}b}+h.c.    \notag \\
&&=\frac{iB}{N_{z}}\sqrt{N^{a+}_{z}(\mathbf{k}_{xy})} \sqrt{N^{b+}_{z}(\mathbf{k}_{xy})}\cdot   \\
&&\cdot\sum\limits_{k_{z}k^{'}_{z}}\frac{\phi^{\dagger}_{\mathbf{k}_{xy}k_{z}a}}{\sqrt{N^{a+}_{z}(\mathbf{k}_{xy})}} f_{k_{z}k^{'}_{z}} s_{0}\otimes\tau_{3} \frac{\phi_{\mathbf{k}_{xy}k^{'}_{z}b}}{\sqrt{N^{b+}_{z}(\mathbf{k}_{xy})}}+h.c. \notag
\end{eqnarray}
In the above summations, $k_{z}$ is within BZ$_{a}$ and $k^{'}_{z}$ is within BZ$_{b}$, different from Eq.(F7). For the $\mathbf{k}_{xy}$ assumed above and small $\Delta_{0}$, $k_{z}$ of the states within BZ$^{\alpha}_{a}$ ($\alpha=\pm$) and BZ$^{\beta}_{b}$ ($\beta=\pm$) are concentrated on very narrow regions centering around $\alpha k_{z0}$ and $\beta k^{'}_{z0}$. So, $f_{k_{z}k^{'}_{z}}$ are concentrated on four values $f_{\alpha k_{z0},\beta k^{'}_{z0}}$ ($\alpha=\pm$, $\beta=\pm$). Accordingly, we can define effective state operators as $\tilde{\phi}_{\mathbf{k}_{xy}a\alpha}=\frac{1}{\sqrt{N^{a+}_{z}(\mathbf{k}_{xy})}}\sum\limits_{k_{z}\in BZ^{\alpha}_{a}}\phi_{\mathbf{k}_{xy}k_{z}a}$ and $\tilde{\phi}_{\mathbf{k}_{xy}b\beta}=\frac{1}{\sqrt{N^{b+}_{z}(\mathbf{k}_{xy})}}\sum\limits_{k_{z}\in BZ^{\beta}_{b}}\phi_{\mathbf{k}_{xy}k_{z}b}$. Then we have
\begin{eqnarray}
&&\bar{\bar{h}}^{z}_{ab}(\mathbf{k}_{xy})\approx\frac{iB\Delta_{0}}{2\pi}\sqrt{\frac{\nu_{a}\nu_{b}} {|v_{z}(\mathbf{k}_{xy},k_{z0})v_{z}(\mathbf{k}_{xy},k^{'}_{z0})|}}\cdot  \notag \\
&&\cdot\sum\limits_{\alpha\beta}\tilde{\phi}^{\dagger}_{\mathbf{k}_{xy}a\alpha}f_{\alpha k_{z0},\beta k^{'}_{z0}} s_{0}\otimes\tau_{3} \tilde{\phi}_{\mathbf{k}_{xy}b\beta}+h.c.
\end{eqnarray}
The factor $\frac{iB\Delta_{0}}{2\pi}\sqrt{\frac{\nu_{a}\nu_{b}} {|v_{z}(\mathbf{k}_{xy},k_{z0})v_{z}(\mathbf{k}_{xy},k^{'}_{z0})|}}f_{\alpha k_{z0},\beta k^{'}_{z0}}$ thus stands as an estimation of average coupling strength between states of orbital $a$ in BZ$^{\alpha}_{a}$ and states of orbital $b$ in BZ$^{\beta}_{b}$. Hereafter, we denote this factor as $C_{\alpha\beta}(\mathbf{k}_{xy})$.

As an example, we estimate the effective coupling strength for $k_{x}=k_{y}=0$, focusing on the two sets of parameters as shown in Figs.2(c) and 2(d). According to former definitions, we have $v_{z}(k_{x}=k_{y}=0,\alpha k_{z0})=2\alpha m_{2}\sin k_{z0}$ and $v^{'}_{z}(k_{x}=k_{y}=0,\beta k^{'}_{z0})=2\beta m_{2}\sin k^{'}_{z0}$. $k_{z0}>0$ satisfies $m_{0}-2m_{2}(1-\cos k_{z0})-\mu=0$, while $k^{'}_{z0}>0$ satisfies $m_{0}-2m_{2}(1-\cos k^{'}_{z0})+\mu=0$. From these relations we have
\begin{eqnarray}
&&|C_{\alpha\beta}(\mathbf{k}_{xy}=\mathbf{0})|=\frac{B\Delta_{0}\sqrt{\nu_{a}\nu_{b}}}{\pi}\cdot  \notag \\ &&\cdot|\frac{\alpha}{\sqrt{(m_{0}+\mu)(\mu-m_{0}-4m_{2})}}   \notag  \\
&&+\frac{\beta}{\sqrt{(\mu-m_{0})(\mu+m_{0}+4m_{2})}}|.
\end{eqnarray}
For parameters of Fig.2(c) of the main text, $m_{0}=-0.7$, $m_{2}=0.5$, $B=0.1$, $\mu=0.2$, and $\Delta_{0}=0.01$, we have
\begin{equation}
|C^{(2c)}_{\alpha\beta}(\mathbf{k}_{xy}=\mathbf{0})|\simeq10^{-4}|4.29\alpha+2.74\beta|\sqrt{\nu_{a}\nu_{b}}.
\end{equation}
For parameters of Fig.2(d) of the main text, $m_{0}=-0.7$, $m_{2}=0.5$, $B=0.3$, $\mu=0.5$, and $\Delta_{0}=0.03$, we have
\begin{equation}
|C^{(2d)}_{\alpha\beta}(\mathbf{k}_{xy}=\mathbf{0})|\simeq10^{-3}|1.95\alpha+7.16\beta|\sqrt{\nu_{a}\nu_{b}}.
\end{equation}

The data in Eqs.(F13) and (F14) can be compared to the numerical results on Figs.2(c) and 2(d). For $\underline{\Delta}_{1}(\mathbf{k})$, both $\tilde{h}_{a}$ defined within BZ$_{a}$ and $\tilde{h}_{b}$ defined within BZ$_{b}$ support SABSs. If we ignore the coupling discussed above, then $\tilde{h}_{a}$ and $\tilde{h}_{b}$ each produces a pair of zero energy SABSs for $k_{x}=k_{y}=0$, which is protected by time reversal symmetry and particle hole symmetry, since $k_{x}=k_{y}=0$ is a time reversal invariant momentum of the surface BZ. The two pairs of zero modes originate separately from coupling among states of orbital $a$ in BZ$_{a}$ and states of orbital $b$ in BZ$_{b}$. Introducing the coupling of $\bar{\bar{h}}^{z}_{ab}(\mathbf{k}_{xy})$ between states of orbital $a$ in BZ$_{a}$ and states of orbital $b$ in BZ$_{b}$ mixes the two pairs of SABSs. For $k_{x}=k_{y}=0$, the coupling strengths are roughly estimated by Eqs.(F13) and (F14), which would turn the two pairs of zero energy SABSs to two pairs (degeneracy protected by time reversal symmetry) of nonzero energy SABSs at two symmetric energies (ensured by particle hole symmetry). From Fig.2(c) of the main text, the numerical results for the excitation energy of the SABSs at $k_{x}=k_{y}=0$ are $\sim\pm0.0021$. A comparison with Eq.(F13) shows that if we take $\nu_{a}$ and $\nu_{b}$ to be between 3 to 4, then Eq.(F13) gives a fairly good estimation of the excitation energy of Fig.2(c) at $k_{x}=k_{y}=0$. From Fig.2(d) of the main text, the numerical results for the excitation energy of the SABSs at $k_{x}=k_{y}=0$ are $\sim\pm0.0068$. A comparison with Eq.(F14) shows that if we take $\nu_{a}$ and $\nu_{b}$ to be about 1, then Eq.(F14) gives a fairly good estimation of the coupling strength and the excitation energy of Fig.2(d) at $k_{x}=k_{y}=0$. Since the required $\nu_{a}$ and $\nu_{b}$ are all in the order of $1$, in agreement with their physical meaning, the present comparison has confirmed the correctness and effectiveness of the picture proposed in the above general analysis.

\emph{\textbf{- Analytical calculation of the SABSs.}}
We now make a realistic analytical calculation of the SABSs for $\underline{\Delta}_{1}$ for parameters similar to Figs.2(c) and 2(d). As was stated in the introductory remarks of this section, we first get two pairs of SABSs by ignoring the coupling between $\tilde{h}_{a}$ and $\tilde{h}_{b}$, then we put back this coupling and get the actual effective model and dispersion of the SABSs.

First consider $\tilde{h}_{a}$, the form of which is as shown in Eq.(3) of the main text. To make possible an analytical analysis, we have restricted to the case of weak pairing field. In addition, we assume that $\sqrt{c^{2}_{x}(\mathbf{k})+c^{2}_{y}(\mathbf{k})+c^{2}_{z}(\mathbf{k})}$ is small as compared to $|m(\mathbf{k})|$. In this limit, the pairing only influences states for which $m(\mathbf{k})-\mu\approx0$. Then, we are justified to replace the $m(\mathbf{k})+\mu$ denominators in Eq.(3) of the main text by $2\mu$. $\tilde{h}_{a}$ thus becomes
\begin{eqnarray}
&&\tilde{h}_{a}\simeq [m-\mu+\frac{c^{2}_{x}+c^{2}_{y}+c^{2}_{z} -\Delta^{2}_{0}}{2\mu}]s_{0}\otimes\tau_{3}   \notag \\
&&-\frac{\Delta_{0}}{\mu} [c_{x}s_{3}\otimes\tau_{1}+c_{y}s_{0}\otimes\tau_{2}-c_{z}s_{1}\otimes\tau_{1}].
\end{eqnarray}
Since $\tilde{h}_{a}$ has particle hole symmetry, we can follow the common practice in deriving the topological surface states of topological insulators.\cite{liu10,hao11}

We first get the pair of zero energy surface modes pertaining to $k_{x}=k_{y}=0$. Since all the relevant parameters for pairing in the BHZ model are close to the BZ center (i.e., $k_{x}=k_{y}=k_{z}=0$), we can expand $\tilde{h}_{a}(k_{x}=k_{y}=0,k_{z})$ into a polynomial in $k_{z}$. By creating a pair of surfaces, $k_{z}$ is not still a good quantum number and should be replaced by $-i\partial_{z}$. Then the pair of zero energy surface modes are obtained by solving
\begin{equation}
\tilde{h}_{a}(k_{x}=k_{y}=0,-i\partial_{z})\psi_{a\alpha}(z)=0,
\end{equation}
where $\alpha=1,2$ labels the two zero modes. To focus on surface localized modes satisfying the boundary condition of $\psi_{a\alpha}(z=0)=\psi_{a\alpha}(z=\pm\infty)=0$ (`$+$' sign applies for the lower surface, while `$-$' sign applies for the upper surface), we set the ansatz for $\psi_{a\alpha}(z)$ as\cite{liu10,hao11}
\begin{equation}
\psi_{a\alpha}(z)=e^{\lambda z}[u_{1},v_{1},u_{2},v_{2}]^{\text{T}},
\end{equation}
where $\lambda$ is a constant to be solved and the superscript `T' means taking the transpose. The four solutions for $\lambda$ are
\begin{eqnarray}
\lambda_{\zeta\eta}&=&\frac{\zeta\Delta_{0}B+\eta\mu\sqrt{4m_{2}(m_{0}-\mu-\frac{\Delta^{2}_{0}}{2\mu})+2B^{2}(\frac{m_{0}}{\mu}-1)}} {2m_{2}\mu+B^{2}}  \notag \\
&=&\frac{\zeta\Delta_{0}B+\eta\mu\sqrt{C_{1}}} {2m_{2}\mu+B^{2}},
\end{eqnarray}
where $\zeta=\pm$ and $\eta=\pm$. For parameters similar to Figs.2(c) and 2(d) of the main text, $C_{1}<0$ and $\sqrt{C_{1}}$ in the numerator is purely imaginary.

Now focus on the surface modes living on the upper surface ($z=0$ surface of a sample occupying $z<0$) of a sample. The relevant solutions for $\lambda$ are $\lambda_{+\eta}$. Substituting Eqs.(F17) and (F18) back into Eq.(F16), we get a pair of zero energy surface modes for $\tilde{h}_{a}(k_{x}=k_{y}=0)$ as
\begin{equation}
\psi_{a\alpha}(z)=C\rho_{\alpha}(e^{\lambda_{++}z}-e^{\lambda_{+-}z}),
\end{equation}
where $\rho_{1}=\frac{1}{\sqrt{2}}[1,0,0,i]^{\text{T}}$ and $\rho_{2}=\frac{1}{\sqrt{2}}[0,1,-i,0]^{\text{T}}$ are the two orthonormal state vectors. The normalization constant $C$ is taken as a positive number
\begin{equation}
C=\frac{1}{\mu}\sqrt{\frac{\Delta_{0}B(\mu^{2}C_{1}-\Delta^{2}_{0}B^{2})}{(2m_{2}\mu+B^{2})C_{1}}},
\end{equation}
where the constant $C_{1}$ is defined in Eq.(F18).

Now consider the SABSs supported by $\tilde{h}_{b}$. Since $\tilde{h}_{b}$ can be obtained from $\tilde{h}_{a}$ by making the substitutions of $m_{\alpha}\rightarrow-m_{\alpha}$ ($\alpha=0,1,2$), $\Delta_{0}\rightarrow-\Delta_{0}$ and $B\rightarrow-B$, the solution of the surface modes can be obtained similar to those for $\tilde{h}_{a}$. Thus we obtain the two zero energy surface modes of $\tilde{h}_{b}$ on the upper surface ($z=0$ surface of a sample occupying $z<0$) for $k_{x}=k_{y}=0$ as
\begin{equation}
\psi_{b\alpha}(z)=D\rho^{'}_{\alpha}(e^{\lambda^{'}_{-+}z}-e^{\lambda^{'}_{--}z}),
\end{equation}
where the two orthonormal state vectors are $\rho^{'}_{1}=\frac{1}{\sqrt{2}}[1,0,0,-i]^{\text{T}}$ and $\rho^{'}_{2}=\frac{1}{\sqrt{2}}[0,1,i,0]^{\text{T}}$, which are complex conjugates of those for $\tilde{h}_{a}$. The four solutions for $\lambda$ are
\begin{equation}
\lambda^{'}_{\zeta\eta}=\lambda_{\zeta\eta}(m_{0}\rightarrow-m_{0},m_{2}\rightarrow-m_{2}) =\frac{\zeta\Delta_{0}B+\eta\mu\sqrt{C_{2}}} {-2m_{2}\mu+B^{2}},
\end{equation}
where $\zeta=\pm$ and $\eta=\pm$. $C_{2}=C_{1}(m_{0}\rightarrow-m_{0},m_{2}\rightarrow-m_{2})$ is negative for parameters similar to those of Figs.2(c) and 2(d). The two roots of $\lambda$ with $\zeta=-$ in Eq.(F22) are taken to define the surface modes localized on the upper surface of a sample, because $2m_{2}\mu>B^{2}$ for parameters in Figs.2(c) and 2(d). The normalization constant $D$ is taken as
\begin{equation}
D=\frac{1}{\mu}\sqrt{\frac{\Delta_{0}B(\mu^{2}C_{2}-\Delta^{2}_{0}B^{2})}{(2m_{2}\mu-B^{2})C_{2}}}.
\end{equation}

Taking $\psi_{a\alpha}(z)$ and $\psi_{b\alpha}(z)$ ($\alpha=1,2$) as four basis states, we can now construct the effective models for the SABSs on the upper $xy$ surface of a sample in the $\underline{\Delta}_{1}$ superconducting phase. First, we calculate the effective model within the subspace of SABSs related to $\tilde{h}_{a}$ and $\tilde{h}_{b}$, still ignoring the coupling between $\tilde{h}_{a}$ and $\tilde{h}_{b}$. By focusing on states close to center of the surface BZ, we can take the terms dependent on $k_{x}$ and $k_{y}$ as perturbations. For $\tilde{h}_{a}$, the perturbation term is
\begin{eqnarray}
\tilde{h}_{axy}=[2m_{1}(3-2\cos\frac{\sqrt{3}}{2}k_{x}\cos\frac{1}{2}k_{y}-\cos k_{y})  \notag \\
+\frac{c^{2}_{y}+c^{2}_{x}}{2\mu}]s_{0}\otimes\tau_{3}-\frac{\Delta_{0}}{\mu}[c_{x}s_{3}\otimes\tau_{1}+c_{y}s_{0}\otimes\tau_{2}].
\end{eqnarray}
Taking $\{\psi_{a1}(z),\psi_{a2}(z)\}$ as the basis, the effective model for the SABSs emerging from $\tilde{h}_{a}$ is
\begin{equation}
\tilde{h}^{surf}_{a}(\mathbf{k}_{xy})=-\frac{\Delta_{0}}{\mu}[c_{x}(\mathbf{k}_{xy})s_{1}+c_{y}(\mathbf{k}_{xy})s_{2}],
\end{equation}
where $s_{1}$ and $s_{2}$ are Pauli matrices acting in the present subspace.

For $\tilde{h}_{b}$, the perturbation term is
\begin{eqnarray}
\tilde{h}_{bxy}=[-2m_{1}(3-2\cos\frac{\sqrt{3}}{2}k_{x}\cos\frac{1}{2}k_{y}-\cos k_{y})  \notag \\
+\frac{c^{2}_{y}+c^{2}_{x}}{2\mu}]s_{0}\otimes\tau_{3}+\frac{\Delta_{0}}{\mu}[c_{x}s_{3}\otimes\tau_{1}+c_{y}s_{0}\otimes\tau_{2}].
\end{eqnarray}
Taking $\{\psi_{b1}(z),\psi_{b2}(z)\}$ as the basis, the effective model for the SABSs emerging from $\tilde{h}_{b}$ is
\begin{equation}
\tilde{h}^{surf}_{b}(\mathbf{k}_{xy})=\frac{\Delta_{0}}{\mu}[c_{x}(\mathbf{k}_{xy})s_{1}+c_{y}(\mathbf{k}_{xy})s_{2}],
\end{equation}
where $s_{1}$ and $s_{2}$ are Pauli matrices acting in the present subspace.

As was explained in the introductory part of this section, the creation of a pair of $xy$ surfaces couples the two pairs of SABSs originating from the decoupled $\tilde{h}_{a}$ and $\tilde{h}_{b}$ together. The dispersion of the actual SABSs close to $k_{x}=k_{y}=0$ are thus obtained after we make a reasonable estimation for the coupling between the two pairs of SABSs solved from $\tilde{h}^{surf}_{a}(\mathbf{k}_{xy})$ and $\tilde{h}^{surf}_{b}(\mathbf{k}_{xy})$. There are two equivalent yet slightly different methods to make this estimation, which we explain in turn.

The first method is to calculate the matrix elements of the original bulk Hamiltonian (with a pair of $xy$ surfaces introduced and thus $k_{z}$ is to be replaced by $-i\partial_{z}$) between the basis $\{\psi_{a1}(z),\psi_{a2}(z)\}$ for $\tilde{h}^{surf}_{a}(\mathbf{k}_{xy})$ and the basis $\{\psi_{b1}(z),\psi_{b2}(z)\}$ for $\tilde{h}^{surf}_{b}(\mathbf{k}_{xy})$. We define the matrix element to be calculated between $\psi_{a\alpha}(z)$ ($\alpha=1,2$) and $\psi_{b\beta}(z)$ ($\beta=1,2$) as $C^{\alpha\beta}_{ab}$. There are four terms in the bulk Hamiltonian of $\underline{\Delta}_{1}$ that are relevant to the coupling between the $a$ and $b$ orbitals, which are
\begin{eqnarray}
&&H_{ab}(\mathbf{k}_{xy},k_{z})=c_{z}(k_{z})\sigma_{2}\otimes s_{0}\otimes\tau_{3}+\Delta_{0}\sigma_{2}\otimes s_{1}\otimes\tau_{1} \notag \\
&&+c_{y}(\mathbf{k}_{xy})\sigma_{1}\otimes s_{1}\otimes\tau_{0}-c_{x}(\mathbf{k}_{xy})\sigma_{1}\otimes s_{2}\otimes\tau_{3}.
\end{eqnarray}
In the four terms of $H_{ab}(\mathbf{k}_{xy},k_{z})$, we have made their dependencies on the wave vectors explicit. Note that the effect of coupling between orbital $a$ and orbital $b$ is already incorporated in obtaining the \emph{bulk} low energy effective models $\tilde{h}_{a}$ within BZ$_{a}$ and $\tilde{h}_{b}$ within BZ$_{b}$. What we are now trying to estimate is the the coupling between states in BZ$_{a}$ and states in BZ$_{b}$ induced by the broken translational invariance along the $z$ direction, so the $k_{z}$ dependent term is the \emph{only} term relevant to our objective. This term is no other than $h^{z}_{ab}$ defined in Eq.(18). For $k_{z}$ very small, we can make the approximation $c_{z}(k_{z})\simeq Bk_{z}$. The breaking of translational invariance along $z$ direction is accompanied by the substitution of $k_{z}\rightarrow -i\partial_{z}$ in $h^{z}_{ab}$. Direct calculation shows that $C^{12}_{ab}=C^{21}_{ab}=0$, and
\begin{eqnarray}
&&C^{11}_{ab}=-C^{22}_{ab}=-BCD(|\lambda_{++}|^{2} +|\lambda^{'}_{-+}|^{2})\cdot  \notag \\
&&\cdot\frac{(\lambda_{++}-\lambda_{+-})(\lambda^{'}_{-+}-\lambda^{'}_{--})}{|(\lambda_{++}+\lambda^{'}_{-+})(\lambda_{+-}+\lambda^{'}_{-+})|^{2}}.
\end{eqnarray}
Thus, in the basis of $\{\psi_{a1}(z),\psi_{a2}(z),\psi_{b1}(z),\psi_{b2}(z)\}$ the complete form of the effective model for the SABSs is
\begin{equation}
h_{SABS}(\mathbf{k}_{xy})=\begin{pmatrix} \tilde{h}^{surf}_{a}(\mathbf{k}_{xy}) & C^{11}_{ab}s_{3} \\
C^{11\ast}_{ab}s_{3} & \tilde{h}^{surf}_{b}(\mathbf{k}_{xy})
\end{pmatrix}.
\end{equation}
Dispersion of the SABSs close to the $k_{x}=k_{y}=0$ point is obtained by diagonalizing the above $4\times4$ matrix, which give four branches of excitations
\begin{equation}
E_{\alpha\beta}(\mathbf{k}_{xy})=\alpha |C^{11}_{ab}|+\beta\frac{\Delta_{0}}{\mu}\sqrt{c_{x}^{2}(\mathbf{k}_{xy})+c_{y}^{2}(\mathbf{k}_{xy})},
\end{equation}
where $\alpha=\pm$ and $\beta=\pm$. This result gives a pair of two fold degenerate states of energy $|C^{11}_{ab}|$ and  $-|C^{11}_{ab}|$ at $k_{x}=k_{y}=0$, which split separately into a pair of linearly dispersive modes for small $\mathbf{k}_{xy}$. This is in qualitative agreement with the dispersions of the SABSs in Figs.2(c) and 2(d). Substituting the parameters for Figs.2(c) and 2(d) to Eq.(F29), we get $|C^{11}_{ab}|=0.0017$ and $|C^{11}_{ab}|=0.0023$, respectively. In comparison to the numerical values of $0.0021$ and $0.0068$, we see that the estimations are in the correct order of magnitude. In particular, the estimation from the analytical calculation for Fig.2(c) is very close to the numerical results, while the estimation for Fig.2(d) is a little inferior. This is because the parameters for Fig.2(c) matches the two conditions of the analytical derivations, small $\Delta_{0}$ and small $\sqrt{c^{2}_{x}(\mathbf{k})+c^{2}_{y}(\mathbf{k})+c^{2}_{z}(\mathbf{k})}$, better than the parameters of Fig.2(d). The approximately constant slope of the analytical dispersion at small $|\mathbf{k}_{xy}|$ for branch $E_{\alpha\beta}(\mathbf{k}_{xy})$ is $\beta A\Delta_{0}/\mu$, which is about 80$\%$ of the numerical value for both Fig.2(c) and Fig.2(d).

The second method of estimating the coupling between the two pairs of SABSs emerging from $\tilde{h}^{surf}_{a}$ and $\tilde{h}^{surf}_{b}$ is to apply Eqs.(F7) and (F9). To begin, we express the basis $\{\psi_{a1}(z),\psi_{a2}(z)\}$ for $\tilde{h}^{surf}_{a}$ and $\{\psi_{b1}(z),\psi_{b2}(z)\}$ for $\tilde{h}^{surf}_{b}$ in lattice representation. Similar to the first part of this section, we denote the numbering of the layers in terms of $n_{z}$. To study the surface states on the upper surface of a sample with $N_{z}$ layers (assuming $N_{z}\gg1$), we set $n_{z}=-1$ as the upmost layer and $n_{z}=-N_{z}$ as the bottom layer. In the case of periodic boundary conditions, $n_{z}=0$ is identified with $n_{z}=-N_{z}$. In the case of open boundary conditions, $n_{z}=0$ is considered as a fictitious layer outside of the sample and just above of the $n_{z}=-1$ layer. Then the surface modes in Eqs.(F19) and (F21) become
\begin{equation}
\psi_{a\alpha}(n_{z})=C\rho_{\alpha}(e^{\lambda_{++}n_{z}}-e^{\lambda_{+-}n_{z}}),
\end{equation}
and
\begin{equation}
\psi_{b\alpha}(n_{z})=D\rho^{'}_{\alpha}(e^{\lambda^{'}_{-+}n_{z}}-e^{\lambda^{'}_{--}n_{z}}),
\end{equation}
where $\alpha=\pm$. The boundary condition of $\psi_{a\alpha}(z=0)=\psi_{b\alpha}(z=0)=0$ are ensured in the form of $\psi_{a\alpha}(n_z=0)=\psi_{b\alpha}(n_z=0)=0$. Note that $\psi_{a\alpha}(n_{z})$ and $\psi_{b\alpha}(n_{z})$ are expressed in the basis of $\phi^{\dagger}_{\mathbf{k}_{xy}=\mathbf{0},n_{z},a}$ (denoted as $\phi^{\dagger}_{n_{z},a}$ hereafter) and $\phi^{\dagger}_{\mathbf{k}_{xy}=\mathbf{0},n_{z},b}$ (denoted as $\phi^{\dagger}_{n_{z},b}$ hereafter) $(\alpha=\pm)$, the four wave functions in Eqs.(F32) and (F33) are written in operator form as
\begin{equation}
\hat{\psi}^{\dagger}_{a\alpha}=C\sum\limits_{n_{z}}\phi^{\dagger}_{n_{z},a}\rho_{\alpha}(e^{\lambda_{++}n_{z}}-e^{\lambda_{+-}n_{z}}),
\end{equation}
and
\begin{equation}
\hat{\psi}^{\dagger}_{b\alpha}=D\sum\limits_{n_{z}}\phi^{\dagger}_{n_{z},b}\rho^{'}_{\alpha}(e^{\lambda^{'}_{-+}n_{z}}-e^{\lambda^{'}_{--}n_{z}}).
\end{equation}
To apply Eqs.(F7) and (F9), we turn the lattice representation of Eqs.(F34) and (F35) to wave vector representation. Since $\tilde{h}_{a}$ and $\tilde{h}_{b}$ are well defined only within BZ$_{a}$ and BZ$_{b}$, $\hat{\psi}^{\dagger}_{a\alpha}$ and $\hat{\psi}^{\dagger}_{a\alpha}$ ($\alpha=\pm$) should also contain states restricted within BZ$_{a}$ and BZ$_{b}$. Recalling the former analysis, there are approximately $2N^{a+}_{z}(\mathbf{k}_{xy}=\mathbf{0})$ (denoted as $2N^{a+}_{z}$ in what follows) states in BZ$_{a}$ and $2N^{b+}_{z}(\mathbf{k}_{xy}=\mathbf{0})$ (denoted as $2N^{b+}_{z}$ in what follows) states in BZ$_{b}$. Similar to the Fourier transformation for states defined in the whole BZ, we can make the following transformations
\begin{equation}
\phi^{\dagger}_{n_{z},a(b)}=\frac{1}{\sqrt{2N^{a(b)+}_{z}}}\sum\limits_{k_{z}\in BZ_{a(b)}}\phi^{\dagger}_{k_{z},a(b)}e^{-ik_{z}n_{z}}.
\end{equation}
Substituting the above transformation to Eqs.(F34) and (F35), we get after finishing the summation over $n_{z}$ from $n_{z}=-N_{z}$ to $n_{z}=-1$
\begin{equation}
\hat{\psi}^{\dagger}_{a\alpha}=\frac{C}{\sqrt{2N^{a+}_{z}}}\sum\limits_{k_{z}\in BZ_{a}}\phi^{\dagger}_{k_{z},a} \frac{\rho_{\alpha}e^{ik_{z}}(e^{\lambda_{+-}}-e^{\lambda_{++}})}{(e^{\lambda_{++}}-e^{ik_{z}})(e^{\lambda_{+-}}-e^{ik_{z}})},
\end{equation}
and
\begin{equation}
\hat{\psi}^{\dagger}_{b\alpha}=\frac{D}{\sqrt{2N^{b+}_{z}}}\sum\limits_{k_{z}\in BZ_{b}}\phi^{\dagger}_{k_{z},b} \frac{\rho^{'}_{\alpha}e^{ik_{z}}(e^{\lambda^{'}_{--}}-e^{\lambda^{'}_{-+}})}{(e^{\lambda^{'}_{-+}}-e^{ik_{z}})(e^{\lambda^{'}_{--}}-e^{ik_{z}})}.
\end{equation}
In deriving the above results, we have exploited the condition of $N_{z}\gg1$, so that $e^{-\lambda_{+\alpha}N_{z}}$ and $e^{-\lambda^{'}_{-\alpha}N_{z}}$ ($\alpha=\pm$) are all essentially zero.

With Eqs.(F37) and (F38) at hand, it is now strait-forward to estimate the coupling between $\{\hat{\psi}_{a\alpha}\}$ and $\{\hat{\psi}_{b\alpha}\}$ in terms of Eqs.(F7) and (F9). Here, we define the matrix element between $\hat{\psi}_{a\alpha}$ ($\alpha=1,2$) and $\hat{\psi}_{b\alpha}$ ($\beta=1,2$) as $D^{\alpha\beta}_{ab}$. It is shown through direct calculation that $D^{11}_{ab}=-D^{22}_{ab}$ and $D^{12}_{ab}=D^{21}_{ab}=0$. So, we get an effective model for the SABSs similar to Eq.(F30), with $C^{11}_{ab}$ replaced by $D^{11}_{ab}$. The final form for $D^{11}_{ab}$ is however more complicated than $C^{11}_{ab}$ and is
\begin{eqnarray}
D^{11}_{ab}&=&\frac{BCD}{4N_{z}\sqrt{N^{a+}_{z}N^{b+}_{z}}}\sum\limits_{k_{z}\in BZ_{a},k^{'}_{z}\in BZ_{b}}(e^{ik_{z}}-e^{ik^{'}_{z}})\cdot  \notag \\ &&\cdot\frac{(e^{\lambda_{+-}}-e^{\lambda_{++}})e^{ik_{z}}}{(e^{\lambda_{++}}-e^{ik_{z}})(e^{\lambda_{+-}}-e^{ik_{z}})}
\cdot  \notag  \\
&&\cdot\frac{(e^{\lambda^{'}_{-+}}-e^{\lambda^{'}_{--}})e^{-ik^{'}_{z}}}{(e^{\lambda^{'}_{-+}}-e^{-ik^{'}_{z}}) (e^{\lambda^{'}_{--}}-e^{-ik^{'}_{z}})}.
\end{eqnarray}
Since the above formula depends on the two parameters $\nu_{a}$ and $\nu_{b}$ defined in Eqs.(F1) and (F2), whose values lack a reliable criterion to determine, we cannot make an easy estimation of its value. But the above derivation has confirmed the equivalence of the two methods, since they give \emph{qualitatively the same} prediction for the dispersion of the SABSs.

\end{appendix}



\begin{references}


\bibitem{schnyder08} A. P. Schnyder, S. Ryu, A. Furusaki, and A. W. W. Ludwig, Phys. Rev. B \textbf{78}, 195125 (2008).

\bibitem{qi09} X. L. Qi, T. L. Hughes, S. Raghu, and S. C. Zhang, Phys. Rev. Lett. \textbf{102}, 187001 (2009).

\bibitem{roy08} R. Roy, arXiv:0803.2868.

\bibitem{sato09} M. Sato, Phys. Rev. B \textbf{79}, 214526 (2009).


\bibitem{qirmp} X. L. Qi and S. C. Zhang, Rev. Mod. Phys. \textbf{83}, 1057 (2011).


\bibitem{hor10} Y. S. Hor, A. J. Williams, J. G. Checkelsky, P. Roushan,
J. Seo, Q. Xu, H. W. Zandbergen, A. Yazdani, N. P. Ong,
and R. J. Cava, Phys. Rev. Lett. \textbf{104}, 057001 (2010).

\bibitem{wray10} L. A. Wray, S. Y. Xu, Y. Xia, Y. S. Hor,
D. Qian, A. V. Fedorov, H. Lin, A. Bansil,
R. J. Cava and M. Z. Hasan, Nature Phys. \textbf{1762}
(2010).

\bibitem{zhang11pa} J. L. Zhang, S. J. Zhang, H. M. Weng, W. Zhang, L. X.
Yang, Q. Q. Liu, S. M. Feng, X. C. Wang, R. C. Yu, L.
Z. Cao, L. Wang, W. G. Yang, H. Z. Liu, W. Y. Zhao, S.
C. Zhang, X. Dai, Z. Fang, C. Q. Jin, Proc. Natl.
Acad. Sci. USA \textbf{108}, 24 (2010).

\bibitem{zhang11pb} C. Zhang, L. Sun, Z. Chen, X. Zhou,
Q. Wu, W. Yi, J. Guo, X. Dong, and Z.
Zhao, Phys. Rev. B \textbf{83}, 140504(R) (2011).


\bibitem{kirshenbaum13} K. Kirshenbaum, P. S. Syers, A. P. Hope, N. P. Butch, J. R. Jeffries, S. T. Weir, J. J. Hamlin, M. B. Maple, Y. K. Vohra, and J. Paglione, Phys. Rev. Lett. \textbf{111}, 087001 (2013).


\bibitem{sasaki12} S. Sasaki, Z. Ren, A. A. Taskin, K. Segawa, L. Fu, and Y. Ando, Phys. Rev. Lett. \textbf{109}, 217004 (2012).


\bibitem{novak13} Mario Novak, Satoshi Sasaki, Markus Kriener, Kouji Segawa, and Yoichi Ando, Phys. Rev. B \textbf{88}, 140502(R) (2013).

\bibitem{fu10} L. Fu and E. Berg, Phys. Rev. Lett. \textbf{105}, 097001
(2010).

\bibitem{hao11} L. Hao and T. K. Lee, Phys. Rev. B \textbf{83}, 134516 (2011).

\bibitem{sasaki11} S. Sasaki, M. Kriener, K. Segawa, K. Yada,
Y. Tanaka, M. Sato, and Y. Ando, Phys.
Rev. Lett. \textbf{107}, 217001 (2011).


\bibitem{hsieh12} T. H. Hsieh and L. Fu, Phys. Rev. Lett. \textbf{108},
107005 (2012).

\bibitem{yamakage12} A. Yamakage, K. Yada, M. Sato, and Y. Tanaka, Phys. Rev. B \textbf{85}, 180509(R) (2012); Y. Tanaka and S. Kashiwaya, Phys. Rev. Lett. \textbf{74}, 3451 (1995); Satoshi Kashiwaya and Yukio Tanaka, Rep. Prog. Phys. \textbf{63}, 1641 (2000).


\bibitem{koren11} G. Koren, T. Kirzhner, E. Lahoud, K. B. Chashka, and
A. Kanigel, Phys. Rev. B \textbf{84}, 224521 (2011).

\bibitem{kirzhner12} T. Kirzhner, E. Lahoud, K. B. Chaska, Z. Salman, and
A. Kanigel, Phys. Rev. B \textbf{86}, 064517 (2012).

\bibitem{chen12} X. Chen, C. Huan, Y. S. Hor, C. A. R. S\'{a} de Melo, and Z. Jiang, arXiv:1210.6054v1.



\bibitem{levy13} N. Levy, T. Zhang, J. Ha, F. Sharifi, A. A. Talin, Y. Kuk, and
J. A. Stroscio, Phys. Rev. Lett. \textbf{110}, 117001 (2013).

\bibitem{hao13} L. Hao, G. L. Wang, T. K. Lee, J. Wang, W. F. Tsai, and Y. H. Yang, Phys. Rev. B \textbf{89}, 214505 (2014).


\bibitem{qi1081} X. L. Qi, T. L. Hughes, and S. C. Zhang, Phys. Rev. B \textbf{81}, 134508 (2010).

\bibitem{sato10} M. Sato, Phys. Rev. B \textbf{81}, 220504(R) (2010).

\bibitem{teo10} J. C. Y. Teo and C. L. Kane, Phys. Rev. B \textbf{82}, 115120 (2010).


\bibitem{basis} S.-K. Yip, Phys. Rev. B \textbf{87}, 104505 (2013); B. Zocher and B. Rosenow, Phys. Rev. B \textbf{87}, 155138 (2013); T. Hashimoto K. Yada, A. Yamakage, M. Sato, and Y. Tanaka, J. Phys. Soc. Jpn. \textbf{82}, 044704 (2013).


\bibitem{qc} Y. Nagai, H. Nakamura, and M. Machida, arXiv:1305.3025; S. Takami, K. Yada, A. Yamakage, M. Sato, and Y. Tanaka, J. Phys. Soc. Jpn. \textbf{83}, 064705 (2014).



\bibitem{qi10} X. L. Qi, T. L. Hughes, and S. C. Zhang, Phys. Rev. B \textbf{82}, 184516 (2010).


\bibitem{kitaev09} A. Kitaev, arXiv:0901.2686v2.


\bibitem{tanaka12}  Yukio Tanaka, Masatoshi Sato, and Naoto Nagaosa, J. Phys. Soc. Jpn. \textbf{81}, 011013 (2012).


\bibitem{nakosai12} Sho Nakosai, Yukio Tanaka, and Naoto Nagaosa, Phys. Rev. Lett. \textbf{108}, 147003 (2012).


\bibitem{michaeli12} Karen Michaeli and Liang Fu, Phys. Rev. Lett. \textbf{109}, 187003 (2012).


\bibitem{zhang13l} Fan Zhang, C. L. Kane, and E. J. Mele, Phys. Rev. Lett. \textbf{111}, 056402 (2013); \emph{ibid} \textbf{111}, 056403 (2013).


\bibitem{bernevig06} B. A. Bernevig, T. L. Hughes, and S. C. Zhang, Science \textbf{314}, 1757 (2006).



\bibitem{wilczeketal} F. Wilczek, Nat. Phys. \textbf{5}, 614 (2009); J. Alicea, Rep. Prog. Phys. \textbf{75}, 076501 (2012); C.W.J. Beenakker, Annu. Rev. Con. Mat. Phys. \textbf{4}, 113 (2013).


\bibitem{balian63} R. Balian and N. R. Werthamer, Phys. Rev. \textbf{131}, 1553 (1963).


\bibitem{salomaa87} M. M. Salomaa and G. E. Volovik, Rev. Mod. Phys. \textbf{59}, 533 (1987); Grigory E. Volovik, \emph{The Universe in a Helium Droplet} (Clarendon Press, Oxford, 2003).


\bibitem{kane05}  C. L. Kane and E. J. Mele, Phys. Rev. Lett.
    \textbf{95}, 146802 (2005); \emph{ibid} \textbf{95}, 226801 (2005).


\bibitem{zhang09} H. Zhang, C. X. Liu, X. L. Qi,
    X. Dai, Z. Fang, and S. C. Zhang, Nature Phys.
\textbf{5}, 438 (2009).



\bibitem{qipt} X. L. Qi and S. C. Zhang, Phys. Today \textbf{63}(1), 33
(2010).



\bibitem{fu09} L. Fu, Phys. Rev. Lett. \textbf{103}, 266801
    (2009).


\bibitem{wang10} Q. H. Wang, D. Wang, and F. C. Zhang,
    Phys. Rev. B \textbf{81}, 035104 (2010).


\bibitem{liu10} C. X. Liu, X. L. Qi, H. J. Zhang,
    X. Dai, Z. Fang, and S. C. Zhang, Phys. Rev. B \textbf{82},
    045122 (2010).


\bibitem{read00} N. Read and D. Green, Phys. Rev. B \textbf{61}, 10267 (2000).



\bibitem{mccann06} We can also work with Green's functions and will get exactly the same low energy effective model. For example, see E. McCann and V. I. Fal'ko, Phys. Rev. Lett. \textbf{96}, 086805 (2006).


\bibitem{hu94} C. R. Hu, Phys. Rev. Lett. \textbf{72}, 1526 (1994).


\end{references}
\end{document}